\documentclass[
12pt,
twocolumn,
]{article}

\newcommand{\bde}{\begin{description}}

\newcommand{\ben}{\begin{enumerate}}
\newcommand{\beq}{\begin{eqnarray}}

\newcommand{\beqn}{\begin{eqnarray*}}

\newcommand{\bmw}{{\mathbf w}}

\newcommand{\bqu}{\begin{quote}}

\newcommand{\bvb}{\begin{verbatim}}

\newcommand{\bw}{{\bf{ w}}}
\newcommand{\bx}{{\bf x}}         

\newcommand{\cf}{{\it cf.}~}

\newcommand{\tty}[1]{\texttt{#1}}

\newcommand{\ea}{\end{eqnarray}}

\newcommand{\ede}{\end{description}}
\newcommand{\een}{\end{enumerate}}
\newcommand{\eeq}{\end{eqnarray}}
\newcommand{\eeqn}{\end{eqnarray*}}
\newcommand{\eg}{{\it e.g.},~}

\newcommand{\equ}{\end{quote}}
\newcommand{\evb}{\end{verbatim}}

\newcommand{\ie}{{\it i.e.},~}

\newcommand{\suppress}[1]{}
\newcommand{\supress}[1]{}

\newcommand{\pmbeg}{\begin{pmatrix}}
\newcommand{\pmend}{\end{pmatrix}}

\renewcommand{\>}{\rangle}

\usepackage{
rotating,graphics,amsmath,amssymb,times,alltt}

\newif\ifpdf
\ifx\pdfoutput\undefined
\pdffalse % we are not running PDFLaTeX
\else
\pdfoutput=1 % we are running PDFLaTeX
\pdftrue \fi
\usepackage{graphicx}

\ifpdf \DeclareGraphicsExtensions{.pdf, .epsf, .tif} \else
\DeclareGraphicsExtensions{.eps } \fi
\date{}
\title{ Discriminative Topological Features Reveal Biological Network Mechanisms}
\author{
Manuel Middendorf$^{1}$,
Etay Ziv$^{2}$,
Carter Adams$^3$,\\
Jen Hom$^4$,
Robin Koytcheff$^4$,
Chaya Levovitz$^5$,\\
Gregory Woods$^3$,
Linda Chen$^6$,
Chris Wiggins$^{7,8}$\\
\small{
$^1$Department of Physics,
$^2$College of Physicians and Surgeons,
$^3$Columbia College,}\\
\small{
$^4$Fu Foundation School of Engineering and Applied Sciences,
$^5$Barnard College,}\\
\small{
$^6$Department of Mathematics,
$^7$Department of Applied Physics and Applied Mathematics,}\\
\small{
$^8$Center for Computational Biology and Bioinformatics;}\\
\small{
Columbia University, New York NY 10027}\\
}
\begin{document}

\maketitle

\newcommand{\nummod}{17~}
\begin{abstract}
    Recent genomic and bioinformatic advances have motivated the development
of numerous
  random network models purporting to describe graphs of biological,
technological,
  and sociological origin.  The success of a model has been evaluated by
  how well it reproduces a few key features of the real-world data,
  such as degree distributions, mean geodesic lengths, and clustering
  coefficients.
  Often pairs of models can reproduce these features with indistinguishable
  fidelity despite being generated by vastly different mechanisms.
  In such cases, these few target features are insufficient to distinguish
  which of the different models best describes real world networks of interest;
  moreover, it is not
  clear a priori that {\it any} of the presently-existing algorithms for network generation
  offers a predictive description of the networks inspiring them.
  To derive discriminative classifiers, we construct a
  mapping from the set of all graphs to a high-dimensional (in principle infinite-dimensional)
  %data
  ``word space.''
This map defines
 %It serves us first to discover new {\it metrics} which are strikingly successful in distinguishing
  %among different network models.  Second, it gives us a means to define
an input space for
  classification schemes which allow us for the first time to state unambiguously which models are
  most descriptive of the networks
%of biological, sociological, and technological interest
they purport to describe.
  Our training sets include networks generated from \nummod
  models either drawn from the literature or introduced in this work,
source code for which is freely available \cite{netclass_website}.
  We anticipate that this new approach to network analysis will be of broad impact to a
  number of communities.
  \end{abstract}

\section{Introduction}
The post-genomic revolution has ushered in an ensemble of novel
crises and opportunities in rethinking molecular biology. The two
principal directions in genomics, sequencing and transcriptome
studies, have brought to light a number of new questions and
forced the development of numerous computational and mathematical
tools for their resolution. The sequencing of whole organisms,
including {\it homo sapiens}, has shown that in fact there are
roughly the same number of genes in men and in mice. Moreover,
much of the coding regions of the chromosomes (the subsequences
which are directly translated into proteins) are highly
homologous. The complexity comes then, not from a larger number of
parts, or more complex parts, but rather through the complexity of
their interactions and interconnections.
% over the top:
% In short, what makes men more than mice is graph theoretic.

Coincident with this biological revolution -- the massive and
unprecedented volume of biological data -- has blossomed a
technological revolution with the popularization and resulting
exponential growth of the Internet. Researchers studying the
topology of the Internet \cite{FFF99} and the World Wide Web
\cite{AJB99} attempted to summarize their topologies via statistical
quantities, primarily the distribution $P(k)$ over nodes of given
connectivity or degree $k$, which it was found, was completely unlike that of a
``random'' or Erdos-Renyi graph \footnote{It will be a question
for historians of science to ponder
  why the Erdos-R\'{e}nyi model of networks was used as the universal
  straw man, rather than the Price model \cite{Price65,Price76},
  inspired by a naturally-occurring graph (the citation graph), which
  gives a power-law degree distribution.}.
%In fact,
Instead,
the distribution
obeyed a power-law 
%was a monotonically (for large $k$) decreasing function
 $P(k)\sim k^{-\gamma}$
for large $k$.
This observation  created a flurry of activity among
mathematicians at the turn of the millennium both in (i) measuring
the degree distributions of innumerable technological,
sociological, and biological graphs (which generically, it turned
out, obeyed such power-law distributions) and (ii) proposing
myriad models of randomly-generated graph topologies which
mimicked these degree distributions (\cf \cite{Newman2003}
for a thorough review).
%needs better wording ``for camera ready''
The success of these latter efforts reveals a conundrum for
mathematical modeling: a metric which is universal (rather
than discriminative) cannot be used for choosing the model which
best describes a network of interest. The question posed is one of {\it classification}, meaning the
construction of an algorithm, based on training data from multiple
classes, which can place data of interest
within one of the classes with small test loss.
%needs better wording

\begin{figure}
  \centerline{
  \includegraphics[width=3.5in]
    {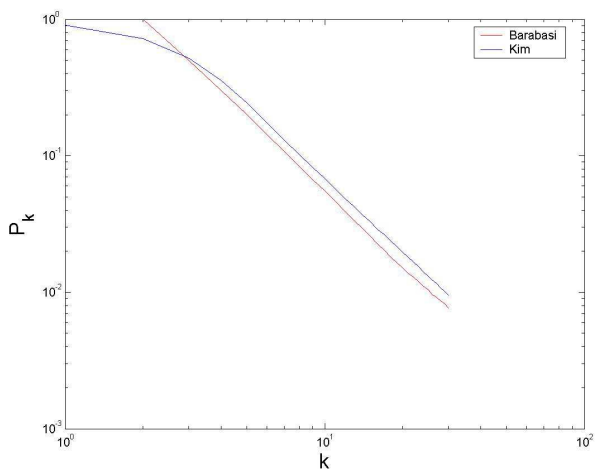}}
\caption{ Ambiguity in network mechanisms: we plot the degree
distribution of two graphs generated using radically different
algorithms. The red line results from an algorithm of the Barabasi
class \cite{AJB99}; the blue from the ``static'' model of Kim et
al \cite{GKK}. The distributions are indistinguishable,
illustrating the insufficiency of degree distributions as a
classifying metric.
%, whereas a single ``word" (see text) is able to distinguish
%hundreds of training examples without error.
} \label{degdist}
\end{figure}

In this paper, we present a natural mapping from a graph to an
infinite-dimensional vector space using simple operations on the
adjacency matrix.
%Furthermore, we use
We then test a number of different classification (including density estimation)
algorithms which prove to be effective in
finding new metrics for classifying real world data sets. We
selected \nummod different models proposed in the literature to
model various properties of naturally occurring networks. Among
them are various biologically inspired graph-generating algorithms
%based on biologically inspired mechanisms
which were put forward to model genetic or protein interaction networks.
To assess
%the value of these models in terms of what they claim to describe,
their value as models of their intended referent,
we classify data sets for the E. coli genetic network,
the C. elegans neural network and the yeast S. cerevisiae protein interaction
network.
We anticipate that this new approach will provide a general
tool of analysis and classification in a broad
diversity of communities.

\begin{figure}[h]
  \centerline{
  \includegraphics[width=2.0 in]{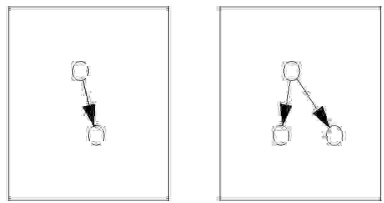}}
  \caption{ The elements of the matrix $ATA$ count these two walks. $TA$ corresponds to one step ``up'' the graph, the following $A$
 to one step ``down''. The last node could be either the same as the starting node as in the first subgraph (accounted for by the diagonal part $DATA$) or a different node as in the second subgraph (accounted for by the non-diagonal part $UATA$). } 
  \label{sumATA}
\end{figure}

The 
input 
space used for classifying graphs was introduced in our
earlier work \cite{matstat} as a technique for finding
statistically significant features and subgraphs in naturally
occurring biological and technological networks.  Given the
adjacency matrix $A$ representing a graph (\ie $A_{ij}=1$ iff
there exists an edge from $j$ to $i$), multiplications of the
matrix count the number of walks from one node to another (\ie
$[A^n]_{ij}$ is the number of unique
walks from $j$ to $i$
in $n$ steps). Note that the adjacency matrix of an undirected
graph is symmetric.
The topological structure of a network is characterized by the number of open and closed walks of given length. Those can be found
by calculating the diagonal or non-diagonal components of the
matrix, respectively. For this we define the projection operation
$D$ such that \beq [D(A)]_{ij}=A_{ij}\delta_{ij} \eeq and its
complement $U=I-D$. (Note that we do not use Einstein's summation convention. Indices $i$ and 
$j$ are not summed over.) We define the primitive alphabet $\{A;T,U,D\}$
as the adjacency matrix $A$ and the operations $T,U,D$
with the transpose operation $T(A)\equiv A^T$ distingushing
walks ``up" the graph from walks "down" the graph. From the {\em letters}
of this
alphabet we can construct {\em words} (a series of operations) of
arbitrary length. A number of redundancies and trivial cases can
be eliminated (for example, the projection operations satisfy
$DU=UD=0$) leading to the operational alphabet
$\{A,AT,AU,AD,AUT\}$. The resulting word is a matrix
representing a set of possible walks generated by the original
graph. An example
is shown in Figure \ref{sumATA}.

Each word determines two relevant statistics
of the network: the number of distinct walks and the number of
distinct pairs of endpoints. These two statistics are determined
by either summing the entries of the matrix (\tty{sum}) or counting the number of
nonzero elements (\tty{nnz}) of the matrix, respectively.
%Thus we map each word onto two integers, $\sum$ and $\cal N$.
Thus the two
operations
\tty{sum} and \tty{nnz}
map words to integers.
This allows us to plot any graph
in a high-dimensional data space:  the coordinates are the integers
resulting from these path-based functionals of the graph's
adjacency matrix.

The coordinates of the infinite-dimensional
data space are given by integer-valued functionals
\beq F(L_1L_2\dots L_nA)\eeq
where each $L_i$ is a letter of the operational alphabet
and $F$ is an operator from the set
$\{{\tty{sum}}, {\tty{sum}} D,$
$ {\tty{sum}} U, {\tty{nnz}}, {\tty{nnz}} D,{\tty{nnz}} U\}$.
We found it necessary only to evaluate
words with $n\leq 4$ (counting all walks up to length 5) to construct low test-loss classifiers.
Therefore, our word space is a
$6\sum_{i=1}^4{5^i}=4680$-dimensional vector space, but since the words
are not linearly independent (\eg ${\tty{sum}} U+{\tty{sum}}
D={\tty{sum}}$), the
dimensionality of the manifold explored
is actually much smaller.
However, we continue to use the full data space since
a particular word, though it may be expressed as a linear combination
of other words,  may be a better discriminator than any of its summands.

In \cite{matstat}, we discuss several possible interpretations of
words, motivated by algorithms for finding subgraphs.
Previously studied metrics can sometimes be
interpreted
in the context of
words.
For example, the \emph{transitivity} of a network can be defined as 3 times
the number of 3-cycles divided by the number of pairs of edges
that are incident on a common vertex.
For a loopless graph (without self-interactions), this can also
be calculated as
a simple expression in word space:
${{\tty{sum}}(DAAA)}/{{\tty{sum}} (UAA)}$. Note that this
expression of transitivity as the quotient of two words implies
separation in two dimensions rather than in one. However, there
are limitations to word space.  For example, a similar measure,
the \emph{clustering coefficient}, defined as the average over all
vertices of the number of 3-cycles containing the vertex divided
by the number of paths of length two centered at that vertex,
cannot be easily expressed in word space because vertices must be
considered individually to compute this quantity. Of course, the
utility of word space is not that it encompasses previously
studied metrics, but that it can elucidate new metrics in an
unbiased, systematic way, as illustrated below.

\section{ Classification Methods}

\subsection{SVMs}
A standard classification algorithm which has been used with great
success in myriad fields is the \emph{support vector machine}, or
SVM \cite{Vapnik}.  This technique constructs a hyperplane in a
high-dimensional feature space separating two classes from each
other.
Linear kernels are used for the analysis presented here; extensions
to appropriate nonlinear kernels are 
possible.

We rely on a freely available C-implementation of SVM-Light
\cite{joachims}, which uses 
a working set selection method to
solve the convex programming problem with Lagrangian \beq
L(\bmw,b)=\frac{1}{2}|\bmw|^2-C\sum_{i=1}^m{\xi_i}\eeq with
$y_i(\bmw\cdot\bx_i+b)\geq 1-\xi_i; i=1,\dots,m$ where
$f(\bx)=\bw\cdot\bx+b$ is the equation of the hyperplane, $\bx_i$
are training examples and $y_i\in \{-1,+1\}$ their class labels. Here, $C$ is
a fixed parameter determining the trade-off between small errors
$\xi_i$ and a large margin ${2}/{|\bmw|}$. 
%\marginpar{maybe rerun with higher C}
We set $C$ to a
default value $(\frac{1}{m}\sum_{i=1}^m{\bx_i^2})^{-1}$.
We observe that 
training and test losses have a
negligible dependence on $C$  since  most test losses
are near or equal to zero even in low-dimensional
projections of the data space.
% -- 
%in short, only a few words are
%generally necessary to distinguish classes of graphs.

\subsection{Robustness}
Our objective is to determine which of
a set of proposed models
most accurately describes a given real data set.
%Our first confidence measure for a classifier is given by low test loss. We
After constructing a classifier enjoying low test loss,
we classify our given real data set to find a `best' model.
However, the real network may lie outside of any
of the sampled distributions of the proposed models in word space.
%In this case, we still obtain a
%prediction of our well-trained classifier, but it is less
%meaningful and thus less useful to us.
In this case we interpret %are read?
our classification as a prediction
of the least erroneous model.

%We propose a second confidence measure for the prediction of
We distinguish between the two cases
%a classifier by using
by noting the following: Consider building a classifier for apples
and grapefruit which is then faced with an orange. The classifier
may then decide that, based on the feature \tty{size} the orange
is an apple. However, based on the feature \tty{taste} the orange
is classified as a grapefruit. That is, if we train our classifier on
different subsets of words and always get the same prediction, the
given real network must come closest to the predicted class based
on any given choice of features we might look at. We therefore
define a {\it robust classifier} as one which consistently
classifies a test datum in the same class, irrespective of the
subset of features chosen. And we measure {\it robustness} as the
ratio of the number of consistent predictions over the total
number of subspace-classifications.

\subsection{Generative Classifiers}
\label{kde}
A generative model, in which one infers
the distribution from which observations are drawn,
allows a quantitative measure of model assignment: the
probability of observing a given word-value given
the model. For a robust classifier, in which assignment
is not sensitively dependent on the set of features chosen,
the conditional probabilities should consistently be greatest for one class.
%SVM lacks the difficulty of distribution estimation and gains
%predictive power by combining multiple features, but lacks the
%quantitative aspect of feature-by-feature generative modeling.

We perform density estimations with Gaussian kernels for each
individual word, allowing calculation of $p(C=c|X_j=x)$, the
probability of being assigned to class $c$ given a particular
value $x$ of word $j$.
% and thus determining which words of a real
%%data network are likely to be generated by a given network model
%and which are not.
By comparing ratios of likelihood values
among the different models, it is therefore possible, for the case of non-robust classifiers, to
determine which of the features of an orange come closest to an
apple and which features come closest to a grapefruit.

We compute the estimated density at a word value $x_0$ from the
training data $x_i$ ($i=1,\dots,m$) as \beq
p(x_0,\lambda)=\frac{1}{m(2\lambda^2\pi)^{1/2}}\sum_{i=1}^m{e^{-\frac{1}{2}(|x_i-x_0|/\lambda)^2}}\eeq
where we optimize the smoothing parameter $\lambda$ by maximizing
the probability of a hold-out set using 5-fold cross-validation.
More precisely, we partition the training examples into 5-folds
$F_i=\{x_{f_i(j)}\}_{j=1\dots N_i}$, where $f_i$ is the set of
indices associated with fold $i$ ($i=1\dots5$) and
$N_i=card(F_i)$. We then maximize \beq
Q(\lambda)=\frac{1}{5}\sum_{i=1}^5\sum_{j=1}^{N_i}{\log{p(x_{f_i(j)},\lambda)}}\eeq
as a function of $\lambda$. In all cases we found that
$Q(\lambda)$ had a well pronounced maximum as long as the data was not
oversampled. Because words can only take integer values,
too many training examples can lead to the situation that
the data take exactly the same values with or without the
hold-out set. In this case, maximizing $Q(\lambda)$ corresponds to
$p(x,\lambda)$ having single peaks around the integer values, so
that $\lambda$ tends to zero. Therefore, we restrict the
number of training examples to $4N_v$, where $N_v$ is the number
of unique integer values taken by the training set. With this
restriction $Q(\lambda)$ showed a well-pronounced maximum at a
non-zero $\lambda$ for all words and models.

\subsection{Word Ranking and Decision Trees}
\label{sec:wordranking}
The simplest scheme to find new metrics which can distinguish
among given models is to take a large number of training
examples for a pair of network models and find the optimal split
between both classes for every word separately. 
%One can 
We then
test every
one-dimensional classifier on a hold-out set and rank words by
lowest test loss. Below we show that this simple approach is
already very successful.

Extending these results,
%%Going one step further
one can ask how many words one needs to
distinguish entire sets of different models,
as estimated by building
%. And one answer could be to build
a multi-class decision tree and
%measure
measuring
its test loss for different numbers of terminal nodes. We use
Matlab's Statistical Toolbox with a binary multi-class cost
function to decide the splitting at each node. To avoid
over-fitting the data, we prune trained trees and select the
subtree with minimal test loss by 10-fold cross-validation.

Additionally, we propose a different approach using decision trees
to find most discriminative words. For every possible model pair
$(i,j)$ for $1\leq i<j\leq N_{mod}$ where  $N_{mod}$ is the total
number of models, we build a binary decision tree, but restricted
so that at every level of each tree the same word has to be used
for all the trees. At every level the best word is chosen
according to the smallest average training loss over all binary
trees. The model is not meant to be a substitution to an ordinary
multi-class decision tree. It merely represents another algorithm
which may be useful to find a fixed number of most discriminative
words, for example for visualization of the distributions in a three-dimensional
subspace.

\suppress{ As an example, the Fig. \ref{degdist} shows

 \begin{figure}[t]
  %\centerline{\includegraphics[width=3.5in] {Ecoli_2w}}
  \centerline{\includegraphics[width=3.5in] {Ecoli_2w}}
  \caption{Superb figure.}
  \label{dataspace}
\end{figure}

Figure \ref{dataspace} (talk about need for generative model)
}

\section{Network Models}

We sample training data for undirected graphs from six growth models, one
scale-free static model \cite{kim}\cite{GKK}\cite{caldarelli}, the Small World model \cite{watts}, and the Erd\"{o}s-R\'{e}nyi model \cite{erdos}.
Among the six growth models two are based on preferential attachment \cite{bianconi}\cite{barabasi},
three on a duplication-mutation mechanism \cite{flammini}\cite{Sole}, and
 one on purely random growth \cite{callaway}.
For directed graphs we similarly train on two preferential attachment models \cite{krapivsky}, 
two static models \cite{grindrod}\cite{higham}\cite{GKK}, three duplication-mutation models \cite{kumar}\cite{vazquez}, and the directed
Erd\"{o}s-R\'{e}nyi model \cite{erdos}. 
More detailed descriptions and source code are available on our website \cite{netclass_website}.

In order to classify real data, we sample training examples of the
given models with a fixed total number of nodes $N_0$, and allow a
small interval $I_M$ of 1-2\% around the total number of edges
$M_0$ of the considered real data set.
All additional model parameters are sampled uniformly over a given range which is specifid by the model's creators in most cases, otherwise can be given reasonable bounds.
 Such a generated graph is accepted if the number of edges
$M$ falls into the specified interval $I_M$ around $M_0$, thereby
creating a distribution of graphs associated to each model which
could  describe the real data set with given $N_0$ and $M_0$.

\section{Results}

We apply our methods to three different real data sets:
the E. coli genetic network \cite{alon}(directed), the S. cerevisiae protein interaction
network \cite{barabasi2}(undirected), and the C. elegans neural
network \cite{White86}(directed).

Each node in E. coli's genetic network represents an operon coding
for a putative transcriptional factor. An edge exists from operon
$i$ to operon $j$ if operon $i$ directly regulates $j$ by binding
to its operator site. This gives a very sparse adjacency matrix with
a total of 423 nodes and 519 edges.

The S. cerevisiae protein interaction network has 2114 nodes and 2203
undirected edges. Its sparseness is therefore comparable to E. coli's
genetic network.

The C. elegans data set represents the organism's fully mapped
neural network.  Here, each node is a neuron and each edge between
two nodes represents a functional, directed connection between two
neurons.  The network consists of 306 neurons and 2359 edges, and
is therefore about 7 times more dense than the other two networks.

We create training data for undirected 
or directed 
models according to the real data set.
All parameters other than
the numbers of nodes and edges were drawn from a uniform
distribution over their range.
We sampled 1000 examples per model for each real data set, trained
a pairwise multi-class SVM on 4/5 of the sampled data and tested
on the 1/5 hold-out set. We determine a prediction by counting
votes for the different classes. 
Table
\ref{svmmain} summarizes the main results.

\begin{table}[h]\begin{center}
\label{table0}
\small{
\begin{tabular}{|l|l|l|l|}
\hline& E. coli&C. elegans&S. cerevisiae\\\hline
$\<L_{tr}\>$&$1.6\%$&$0.5\%$&$2.1\%$\\\hline
$\<L_{tst}\>$&$1.6\%$&$0.5\%$&$1.8\%$\\\hline
$\<N_{sv}\>$&109&51&106\\\hline
 Winner&Kumar&MZ&Sole\\\hline
 Robustness&1.0&.97&0.64\\\hline
\end{tabular}
}
\end{center}
\caption{ Results of multi-class SVM. $\<L_{tr}\>$ is the
empirical training loss averaged over all pairwise
classifiers, $\<L_{tst}\>$ is the averaged empirical test loss.  $\<N_{sv}\>$ is
the average number of support vectors. The winner is the model
that got the highest number of votes when classifying the given
real data set.} \label{svmmain}\end{table}

All three classifiers show very low test loss and two of them a
very high robustness. The average number of support vectors is
relatively small. 
%The relatively small average number of support
%vectors indicates an classification task. 
Indeed, some
pairwise classifiers had as few as three support vectors
and more than half of them had zero test loss. All of this suggests
the existence of a small subset of words which can
distinguish among 
%all 
most of these models.

%It is interesting to note that 
The predicted models Kumar, MZ, and
Sole are based on very similar mechanisms of duplication and
mutation. The model by Kumar {\em et al} was originally meant to
explain various properties of the WWW. It is based on a
duplication mechanism, where at every time step a prototype for
the newly introduced node is chosen at random, and connected to
the prototype's neighbors or other randomly chosen nodes with
probability $p$. It is therefore built on an imperfect copying
mechanism which can also be interpreted as duplication-mutation,
 often evoked when considering genetic and
protein-interaction networks. Sole is based on the same idea, but
allows two free parameters, a probability controlling the number of
edges copied and a probability controlling the number of
random edges created. MZ is essentially a directed version of
Sole.  Moreover, we observe that none of the
preferential attachment models came close to being a predicted model for
one of our biological networks
even though they, and other preferential attachment models in the literature,
were created to explain power-law degree distributions.
The duplication-mutation scheme arises as the more successful one.

Kumar and MZ were classified with almost perfect robustness
against 500-dimensional subspace sampling. With 26 different
choices of subspaces, E. coli was always classified as Kumar. We
therefore assess with high confidence that Kumar and MZ come
closest to modeling E. coli and C. elegans, respectively. In the
case of Sole and the S. cerevisiae protein network we observed
fluctuations in the assignment to the best model. 3 out of 22
times S. cerevisiae was classified as Vazquez (duplication-mutation)
, other times as Barabasi (preferential attachment),
Klemm (duplication-mutation), Kim (scale-free static) or Flammini (duplication-mutation) depending on the subset of words chosen.
This clearly indicates that different features support different
models. Therefore the confidence in classifying S. cerevisiae to be Sole
is limited.

\begin{figure}[h]
  \centerline{\includegraphics[width=3.3in] {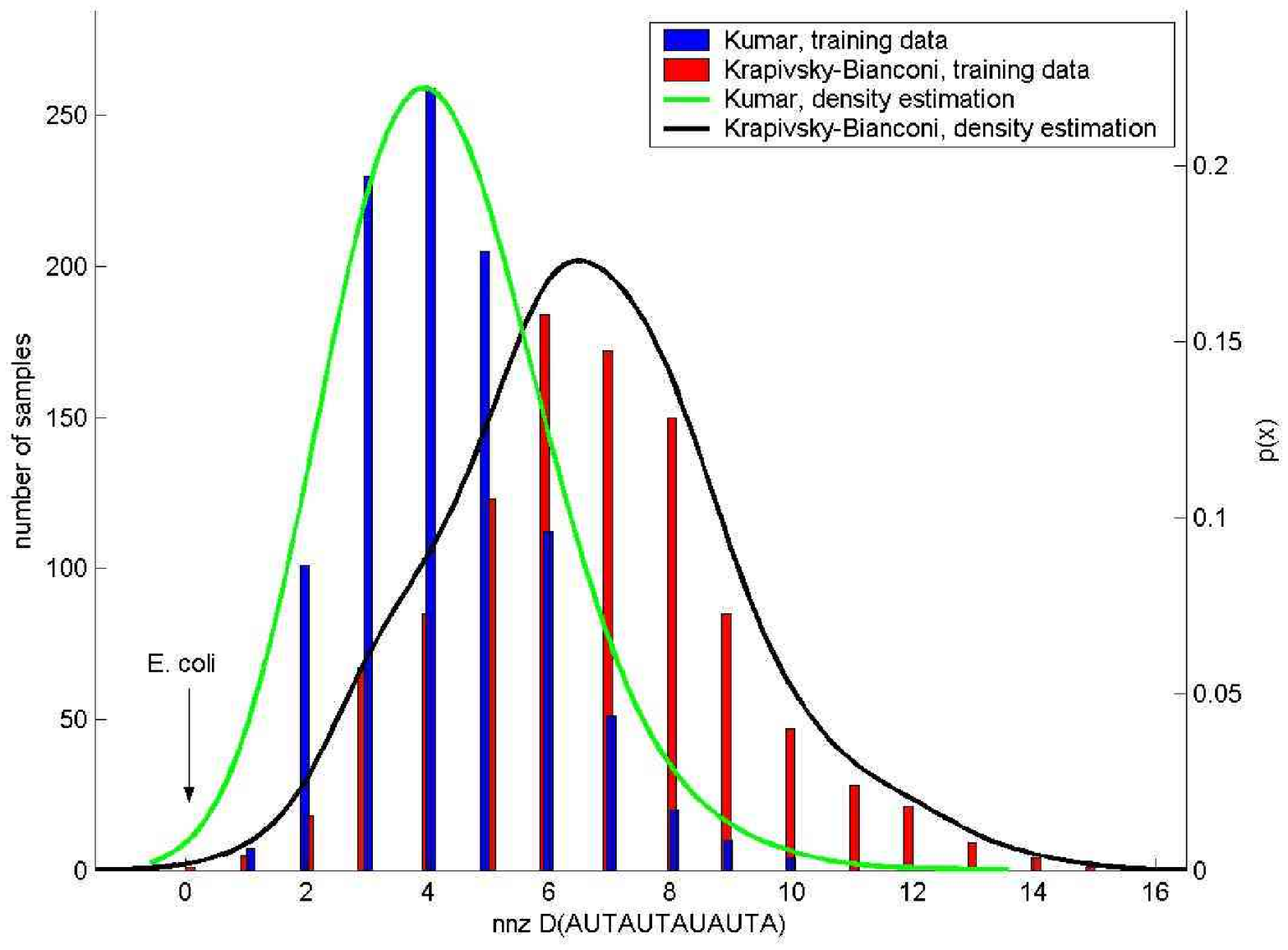}}
  \caption{Kernel Density Estimation of {\tty nnz}~$D(AUTAUTAUAUTA)$ for two
models of E. coli (Kumar and Krapivsky-Bianconi). Log-Likelihoods:
$\log(p_{kumar})=-4.22, \log(p_{krap-bianc})=-12.0$.}
  \label{fig:coli_kde_max}
\end{figure}

\suppress{
\begin{figure}[h]
  \centerline{\includegraphics[width=3.3in] {coli_kde_min_yy}}
  \caption{Kernel Density Estimation of $nnz U(AUTA)$ for two top-scoring
models of C. elegans (Kumar and Krapivsky-Bianconi). Likelihoods:
$\log(p_{kumar})<-744, \log(p_{krap-bianc})=-7.14$.}
  \label{fig:coli_kde_min}
\end{figure}
\begin{figure}[h]
  \centerline{\includegraphics[width=3.3in] {celeg_kde_max_yy}}
  \caption{Kernel Density Estimation of $nnz D(AUTAUTAUA)$ for two top-scoring
models for C. elegans (Middendorf-Ziv and Grindrod).
Log-Likelihoods: $\log(p_{mz})=-4.39, \log(p_{grind})<-744$.}
  \label{fig:celegans_kde_max}
\end{figure}
}
\begin{figure}[h]
  \centerline{\includegraphics[width=3.3in] {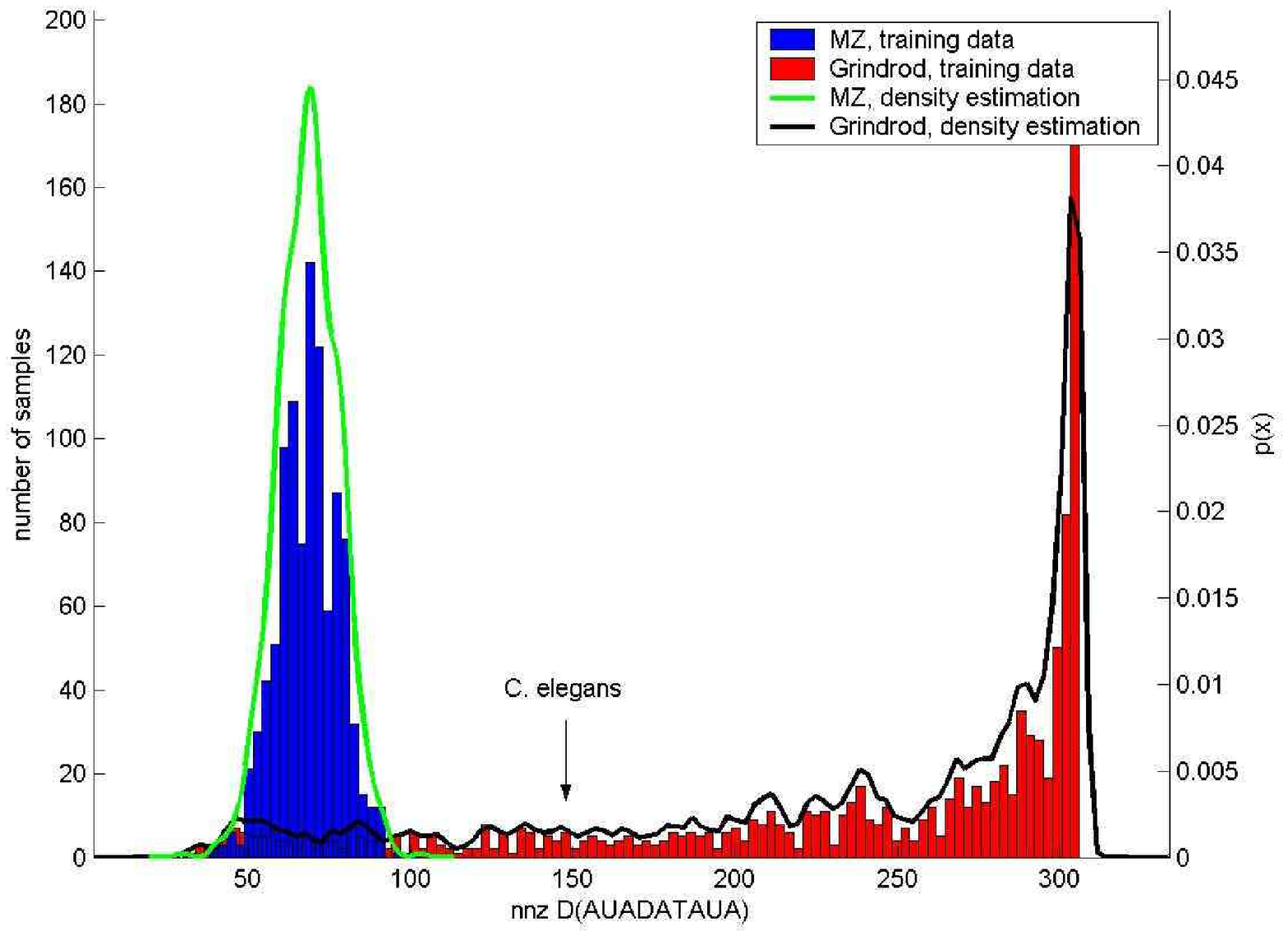}}
  \caption{Kernel Density Estimation of {\tty nnz}~$D(AUADATAUA)$ for two top-
scoring models of C. elegans (Middendorf-Ziv and Grindrod).
Log-Likelihoods:  $\log(p_{mz})=-376, \log(p_{grind})=-6.23$.}
  \label{fig:celegans_kde_min}
\end{figure}
\suppress{
\begin{figure}[h]
  \centerline{\includegraphics[width=3.3in] {yeast_kde_min_yy}}
  \caption{Kernel Density Estimation of $nnz D(AAA)$ for two
models of S. cerevisiae (Sole and Vazquez). Log-Likelihoods:
$\log(p_{vazquez})<-744, \log(p_{sole})=-12.0$. S. cerevisiae has
word value of 1870 in this case and lies far away from any of the
two distributions. Clearly there exist features which neither Sole nor Vazquez
correctly model.}
  \label{fig:yeast_kde_max}
\end{figure}
\begin{figure}[h]
  \centerline{\includegraphics[width=3.3in] {yeast_kde_max_yy}}
  \caption{Kernel Density Estimation of $nnz U(AADAAA)$ for two top-
scoring models of S. cerevisiae (Sole and Vazquez).
Log-Likelihoods: $log(p_{sole})=-407, log(p_{vazquez})=-433$;
$x_{S. cerevisiae}=64658$}
  \label{fig:yeast_kde_min}
\end{figure}
}

The preference of individual words for individual models is
investigated using kernel density estimation \ref{kde} by finding words which maximize $p_i(x_0)/p_j(x_0)$
for two different models ($i$ and $j$) at a word value of the real
data set $x_0$. Figure 
\ref{fig:celegans_kde_min} shows the sampled distribution and
estimated density for the word which extremely
 disfavors the winning model over its follower. The opposite case is shown in
\ref{fig:coli_kde_max} for E. coli,
where the word supports the winning model
and disfavors its follower. More specifically we are able to
verify that most of the words of E. coli are most likely to be
generated by Kumar. Indeed, out of 1897 words taking at least 2
integer values for all of the models (density estimation for a
single value is not meaningful), the estimated density at the E.
coli word value was highest for Kumar in 1297 cases, for
Krapivsky-Bianconi in 535 cases and for Krapivsky in only 65
cases. 
\begin{figure}
\centerline{\includegraphics[width=2.5in] {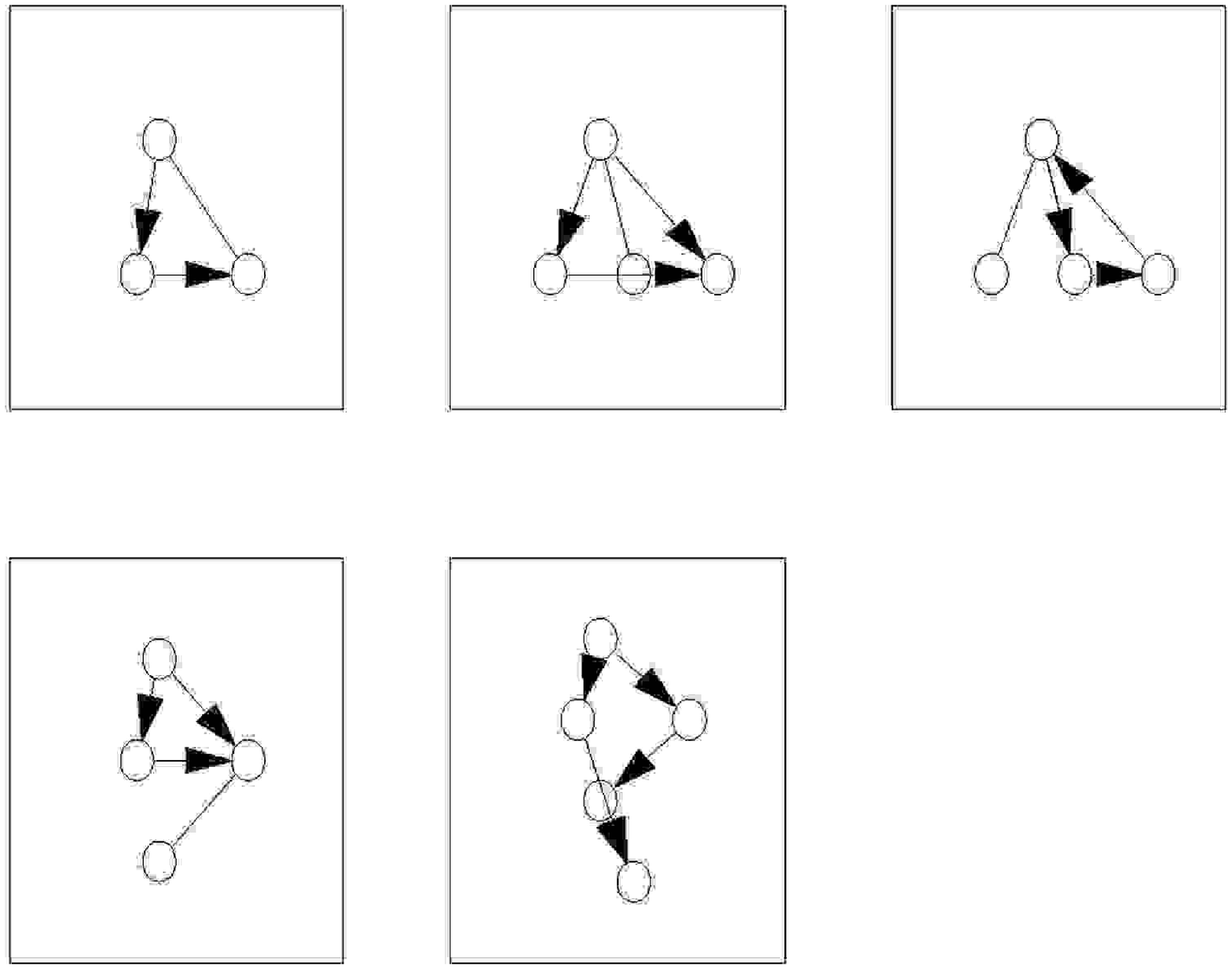}}
  \caption{Subgraphs associated with the word {\tty nnz}$DAUTAUTAUAUTA$. The word has a non-zero value iff at least one of these subgraphs occurs in the network 
  }
  \label{nnzetc}
\end{figure}

Figure \ref{fig:coli_kde_max} shows the distributions for the word {\tty nnz}$DAUTAUTAUAUTA$ which had a
maximum ratio of probability density of Kumar over the one of
Krapivsky-Bianconi at the E. coli position. E. coli in fact has a
zero word count meaning that none of the associated subgraphs shown in Figure \ref{nnzetc} 
actually occur in E. coli. Four of those subgraphs have a mutual edge which is absent in the E. coli network and also impossible to generate in a Kumar graph. 
Krapivsky-Bianconi graphs allow for mutual edges which could be one of the reasons for a higher count in this word. Another source might be that the fifth subgraph showing a higher order feed-forward loop is more probable to be generated in a Krapivsky-Bianconi graph than in a Kumar graph. This subgraph also has to be absent in the E. coli network since it gives a zero word value, showing that the Kumar and the Krapivsky-Bianconi models have both a tendence to give rise to a topological structure that does not exist in E. coli. This analysis gives an example of how these findings are useful in refining network models and in deepening our understanding of real networks.
For further discussions refer to our website \cite{netclass_website}

\begin{figure}[h]
  \centerline{\includegraphics[width=3.3in] {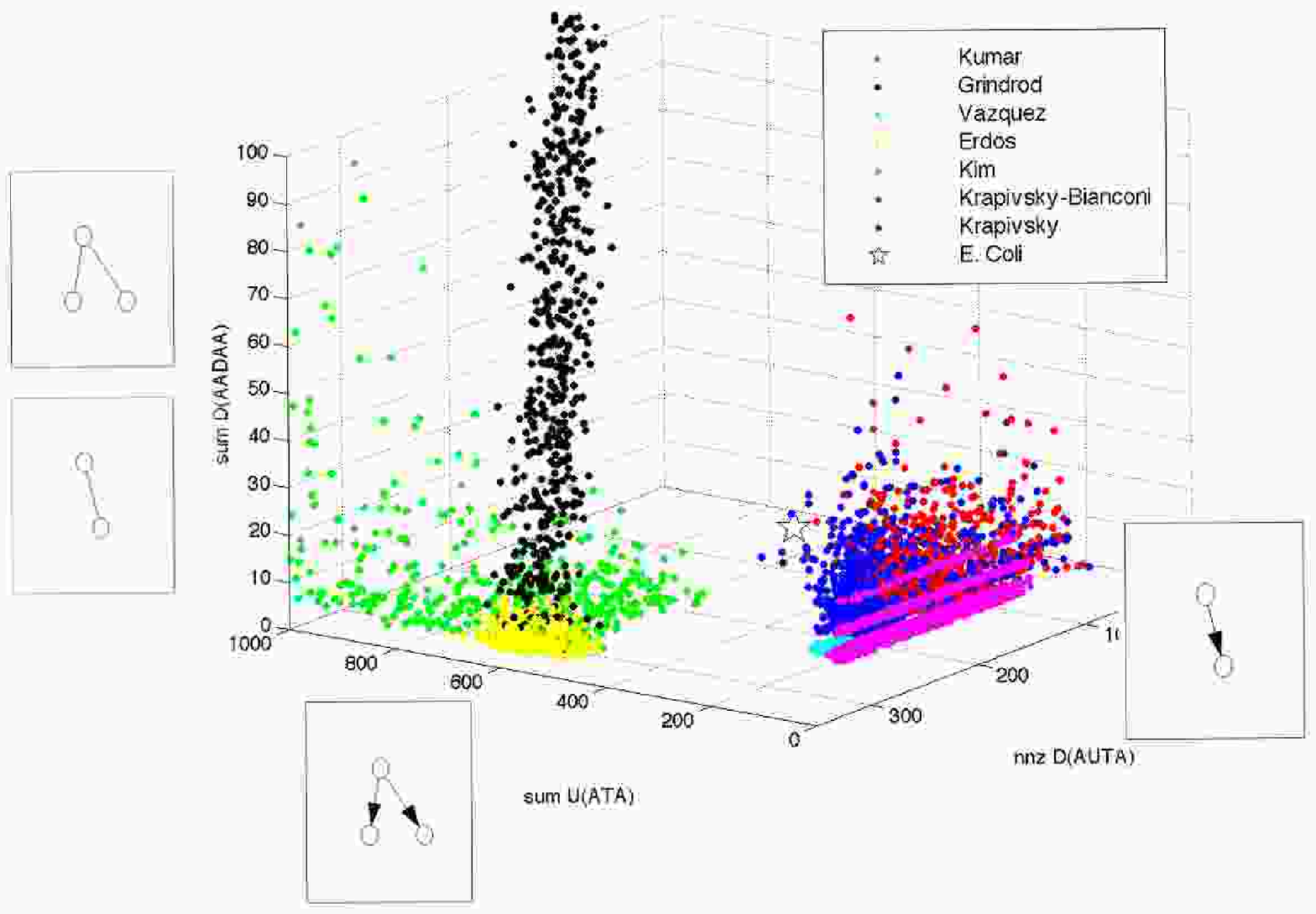}}
  \caption{E. coli and seven directed models. The distributions in word space are shown
for a projection onto the subspace of the three most discriminatve words. 
Subgraphs associated with every word are also shown.}
  \label{ecoli}
\end{figure}
\suppress{
 \begin{figure}[t]
  \centerline{\includegraphics[width=3.3in] {celegans3}}
  \caption{C. elegans and seven directed models. The distributions in word space are sho
wn for a projection onto the subspace of the three most discriminative words. Subgraph
s associated with every word are also shown.}
  \label{scatterceleg}
\end{figure}
}
 \begin{figure}[t]
  \centerline{\includegraphics[width=3.3in] {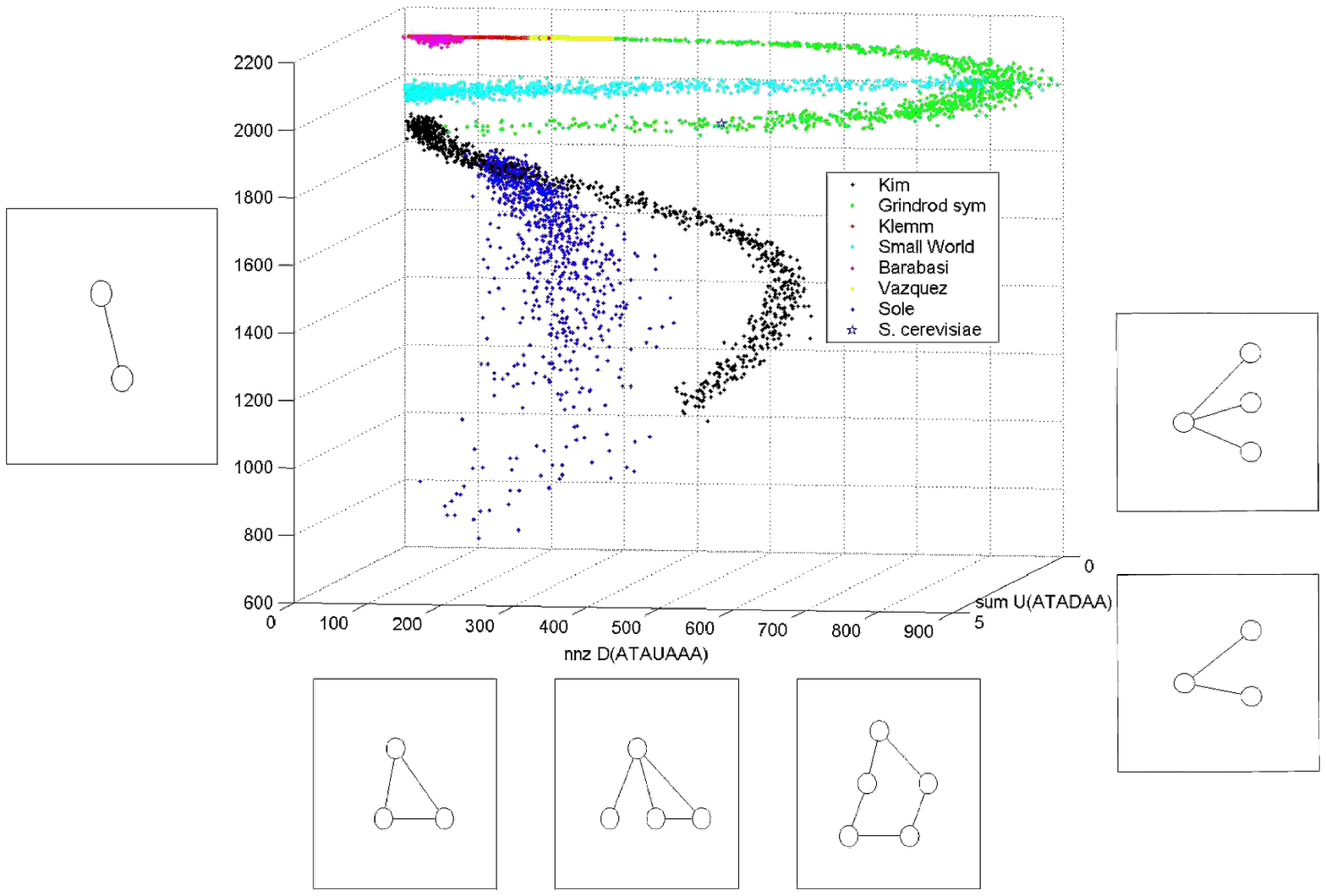}}
  \caption{S. cerevisiae and 7 undirected models. The distributions in word space are 
shown for a projection onto the subspace of the three most discriminative words.
Subgraphs associated with every word are also shown.}
  \label{yeast}
\end{figure}

The SVM results suggest that one may only need  a small subset
of words to be able to separate most of the models with almost zero
test loss. The simplest approach to find such a
subset is to look at every word for a given pair of models and
compute the best split, then ranking words by lowest training loss. 
%Finding the words which minimize test loss
%gives the Tables \ref{ecolisingle}, \ref{celegsingle} and
%\ref{yeastsingle}.
We find that among the most discriminative words some occur very
often such as {\tty nnz}$AA$ or {\tty nnz}$ATA$, which count the pairs of edges
attached to the same vertex and either pointing in the same
direction or pointing away from each other, respectively.
Other frequent words include {\tty nnz}$DAA$, {\tty nnz}$DATA$ and {\tty sum}$UATA$. 
%Other sets of subgraphs corresponding to discriminative words are shown in Figure %\ref{fig:discrim_word}.
A striking feature of this single-word analysis is that the test loss associated with
simple one-dimensional classifiers are comparable to the
SVM test loss
%calculated in Table \ref{table0}. 
%\footnote{For
%Krapivsky-Bianconi vs. Krapivsky the test and training loss are
%actually higher for the SVM than for the single word split. This
%startling fact is very likely due to a low choice of C.}. 
confirming that most pairs of models are separable with
only a few words.
To consider all of the models at once and not just in pairs we
apply both tree algorithms described in \ref{sec:wordranking} to
all three data sets. Figures 
%\ref{scatterceleg}, 
\ref{ecoli} and
\ref{yeast} show scatter-plots of the training data using the most discriminative
three words.
% found by the pairwise binary trees. 
Taking those three words the average training-loss over all pairs of models is 1.7\%, 0.8\% and 0.2\% for the E. coli, C. elegans and S. cerevisiae training data, respectively.
\suppress{
\begin{figure}[h]
  \centerline{\includegraphics[width=3.3in] {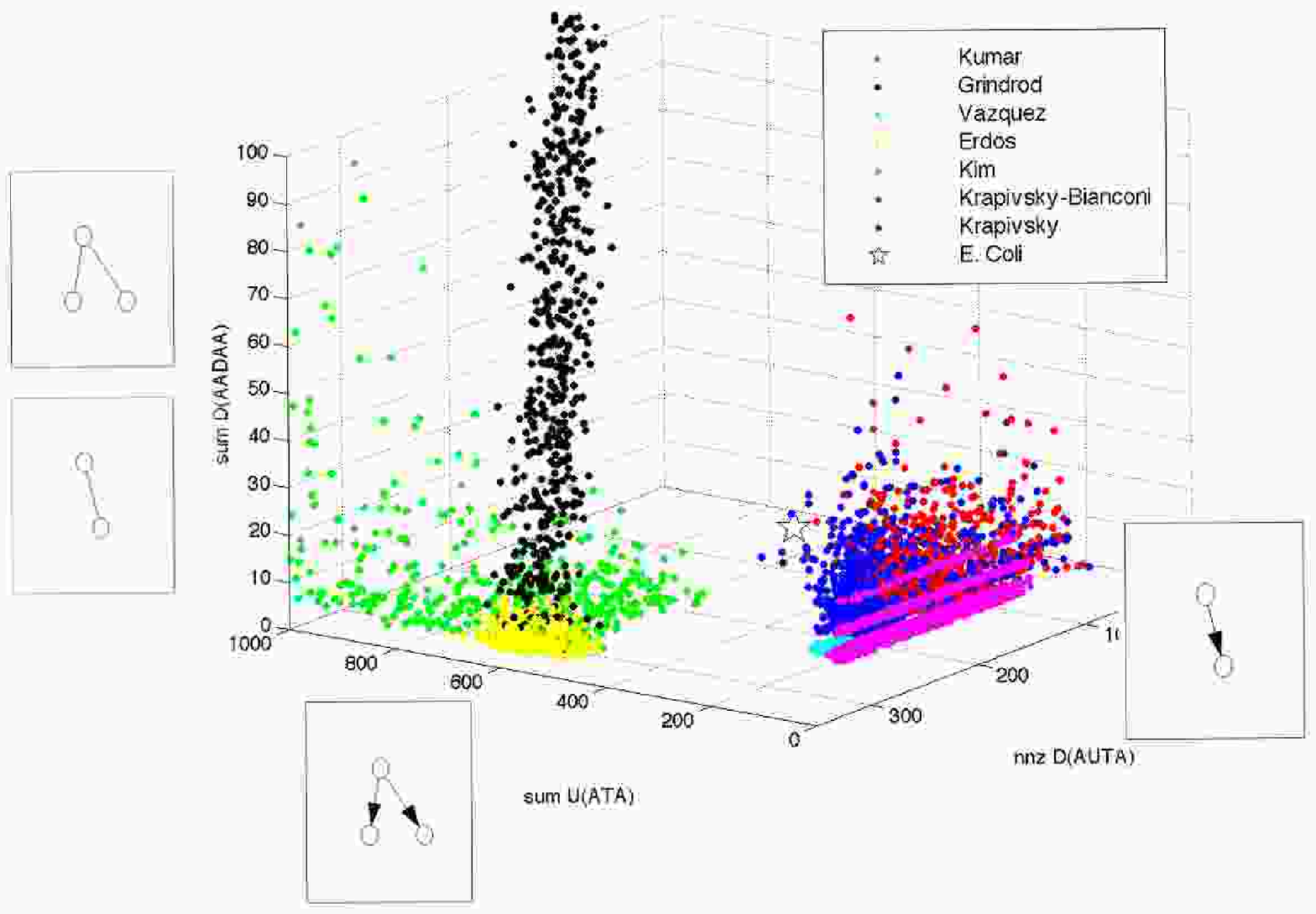}}
  \caption{E. coli and 7 directed models. The distributions in word space are shown for a projection onto a subspace of the three most discriminatve words. Subgraphs associated with every word are also shown.}
  \label{ecoli}
\end{figure}
\suppress{
 \begin{figure}[t]
  \centerline{\includegraphics[width=3.3in] {celegans3}}
  \caption{C. elegans and 7 directed models. The distributions in word space are shown for a projection onto a subspace of the three most discriminative words. Subgraphs associated with every word are also shown.}
  \label{scatterceleg}
\end{figure}
}
 \begin{figure}[t]
  \centerline{\includegraphics[width=3.5in] {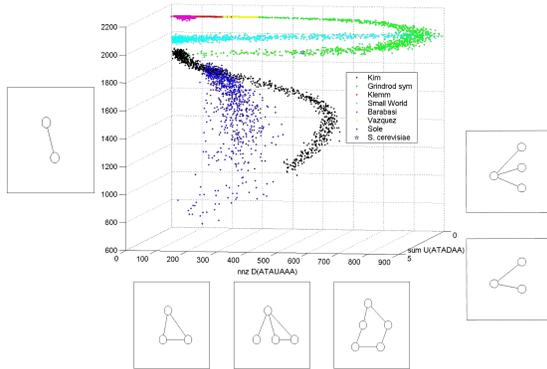}}
  \caption{S. cerevisiae and 7 undirected models. The distributions in word space are
shown for projection onto a subspace of the three most discriminative words.
Subgraphs associated with every word are also shown.}
  \label{yeast}
\end{figure}
}

\section{Conclusions}

It is not surprising that models with different mechanisms are
distinguishable; however, the fact that these models have not
been separated in a systematic manner to date points to the inadequacy of
current
metrics popular in the network theory community.
%Alone, degree
%distributions have not proven themselves to be useful metrics.
We have shown
 that a systematic enumeration of countably infinite features of
graphs can be successfully used to find new
metrics which are highly efficient in separating
various kinds of models.
Furthermore, they allow us to define a high-dimensional
input space for classification algorithms which for the
first time are able to decide which of a given set of models most
accurately describes three exemplary biological networks.

\section{Acknowledgments}
It is a pleasure to acknowledge useful conversations with C.
Leslie, D. Watts, and P. Ginsparg.  We also acknowledge the
generous support of NSF VIGRE grant DMS-98-10750, NSF
ECS-03-32479, and the organizers of the LANL CNLS 2003 meeting and the COSIN
midterm meeting 2003.
%\bibliographystyle{plain}
%\bibliography{netclass}
\appendix
%\multicolumn{1}
%\onecolumn

\vfil\eject
%\begin{widetext}
\onecolumn
\section{Supplementary tables}
\small{
\begin{table}[h]
\begin{center}
\small{
\begin{tabular}{|l|l|l|}
\hline
 \emph{Name}& \emph{Fundamental
 Mechanism}&\emph{References}\\\hline\hline

Bianconi&\parbox{8cm}{Growth model with a probability of attaching
to an existing node $p\sim \eta_i k_i$, where $\eta_i$ is a
fitness parameter. Here we use a random fitness landscape, where
$\eta$ is drawn from a uniform distribution in
$(0,1)$}&\cite{bianconi}\\\hline

Callaway& \parbox{8cm}{Growth model adding one node and several
edges between randomly chosen existing nodes (not necessarily the
newly introduced one) at every time step.
%Our code can either allow for the creation of self-loops or not.
}&
\cite{callaway}\\\hline

Kim & \parbox{8cm}{A ``static" model giving rise to a scale-free
network. Edges are created between nodes chosen with a probability
$p\sim i^{-\alpha}$ where $i$ is the label of the node and
$\alpha$ a constant parameter in $(0,1)$.} &
\cite{kim},\cite{GKK},\cite{caldarelli}\\\hline

Erdos & \parbox{8cm}{Undirected random graph.
%We can either allowfor self-loops or not.
}&\cite{erdos} \\\hline

Flammini & \parbox{8cm}{Growing graph based on duplication
modeling protein interactions. At every time step a prototype is
chosen randomly. With probability $q$ edges of the prototype are
copied. With probability $p$ an edge to the prototype is created.}
& \cite{flammini}\\\hline

Klemm & \parbox{8cm}{Growing graph using sets of active and
inactive nodes to model citation networks.}  & \cite{klemm},
\cite{klemm2}\\\hline

Small World& \parbox{8cm}{Interpolation between a regular lattice
and a random graph. We replace edges in the regular lattice by
random ones.
%We can allow for self-loops or not.
} &
\cite{watts}\\\hline

Barabasi &\parbox{8cm}{Growing graph with a probability of
attaching to an existing node $p\sim k_i$. (``Bianconi" with
$\eta_i=1$ for all $i$) }&\cite{barabasi}\\\hline

Sole &\parbox{8cm}{Growing graph initialized with a 5-ring
substrate. At every time step a new node is added and a prototype
is chosen at random. The prototype's edges are copied with a
probability $p$. Furthermore, random nodes are connected to the
newly introduced node with probability $q/N$, where $p$ and $q$
are given parameters in $(0,1)$ and $N$ is the number of total
nodes at the considered time step.} & \cite{Sole}\\\hline
\end{tabular}
}
\end{center}\caption{Undirected Network Models. $k_i$ is the degree of the $i$-th node. }
\label{model_undirected}
\end{table}

\begin{table}[h]\begin{center}\begin{tabular}{|p{.8 in}|l|l|}
\hline\hline Name& Fundamental Mechanism&References\\\hline

Kim\footnote{We give the same name to both the undirected and the
directed version. It will be clear from the context which one is
meant} &
\parbox{8cm}{Directed version of ``Kim". A ``static" model giving
rise to a scale-free network. Edges are created between nodes
chosen with probabilites $p\sim i^\alpha_{in}$ and $q\sim
j^\alpha_{out}$ where $\alpha_{in}$ and $\alpha_{out}$ are fixed
parameters chosen in $(0,1)$ and $i$($j$) is the label of the
$i$-th ($j$-th) node }&\cite{GKK}\\\hline

Erdos&  \parbox{8cm}{Directed random graph.
%We can either allowfor self-loops or not.
}&\cite{erdos}\\\hline

Grindrod & \parbox{8cm}{Static graph. Edges are created between
nodes $i$, $j$ with probability $p= b\lambda^{|i-j|}$, where $b$
and $\lambda$ are fixed parameters.
}&\cite{higham},\cite{grindrod}\\\hline

Krapivksy& \parbox{8cm}{Growing graph modeling the WWW. At every
time step either a new edge, or a new node with an edge, are
created. Nodes to connect are chosen with probability $p\sim
k_{i,in}+a$ and $q\sim k_{j,out}+b$ based on preferential
attachment with fixed real-valued offsets $a$ and $b$.
%We can either allow for self-loop or not.
}& \cite{krapivsky}\\\hline

Krapivsky-Bianconi& \parbox{8cm}{Extension of ``Krapivsky" using a
random fitness landscape multiplying the probabilities for
preferential attachment. It is the directed analog of ``Bianconi"
being an extension to ``Barabasi". }& (original)\\\hline

Kumar & \parbox{8cm}{Growing graph based on a copying mechanism to
model the WWW.   At every time step a prototype $P$ is chosen at
random. Then for every edge connected to $P$, with probability $p$
an edge between the newly introduced node and $P$'s neighbor is
created, and with probability $(1-p)$ an edge between the new node
and a randomly chosen other node is created. }&
\cite{kumar}\\\hline

Middendorf-Ziv (MZ)& \parbox{8cm}{Growing directed graph modeling
biological network dynamics.  A prototype is chosen at random and
duplicated.  The prototype or progenitor node has edges pruned
with probability $\beta$ and edges added with probability
$\alpha\ll\beta$.  Based loosely on the undirected protein network
model of Sole et al. \cite{Sole}.}& original\\\hline

Vazquez & \parbox{8cm}{Growth model based on a recursive `copying'
mechanism, continuing to 2nd nearest neighbors, 3rd nearest
neighbors etc. The authors call it a `random walk' mechanism.} &
\cite{vazquez}\\\hline

\end{tabular}\end{center}\caption{Directed Network Models. $k_{i,in}$ ($k_{i,out}$) is the in-(out-)degree
 of the $i$-th node. }\label{model_directed}\end{table}

\begin{sidewaystable}[h]\begin{center}
\small{
\begin{tabular}{|p{.8 in}|l|l|l|l|l|l|l|l|l|}
\hline
&votes&Kumar&Krapivsky-Bianconi&Krapivsky&Kim&Vazquez&Erdos&Grindrod&MZ\\\hline
Kumar &7/7&&$f(\bx)=1.48$&$f(\bx)=2.32$&$f(\bx)=2.80$&$f(\bx)=1.12$&$f(\bx)=3.58$&$f(\bx)=3.11$&$f(\bx)=1.26$\\
&&&$L_{tst}=5.3$\%&$L_{tst}=4.5$\%&$L_{tst}=0.8$\%&$L_{tst}=0.0$\%&$L_{tst}=0.0$\%&$L_{tst}=0.0$\%&$L_{tst}=0.0$\%\\
&&&$L_{tr}=4.4$\%&$L_{tr}=3.2$\%&$L_{tr}=0.7$\%&$L_{tr}=0.0$\%&$L_{tr}=0.0$\%&$L_{tr}=0.0$\%&$L_{tr}=0.0$\%\\
&&&$N_{sv}$=139&$N_{sv}$=122&$N_{sv}$=194&$N_{sv}$=9&$N_{sv}$=10&$N_{sv}$=9&$N_{sv}$=9\\\hline
Krapivsky-Bianconi &6/7&$f(\bx)=-1.48$&&$f(\bx)=2.44$&$f(\bx)=2.49$&$f(\bx)=1.01$&$f(\bx)=2.33$&$f(\bx)=2.30$&$f(\bx)=1.64$\\
&&$L_{tst}=5.3$\%&&$L_{tst}=32.8$\%&$L_{tst}=0.8$\%&$L_{tst}=0.0$\%&$L_{tst}=0.0$\%&$L_{tst}=0.0$\%&$L_{tst}=0.0$\%\\
&&$L_{tr}=4.4$\%&&$L_{tr}=31.3$\%&$L_{tr}=0.9$\%&$L_{tr}=0.0$\%&$L_{tr}=0.0$\%&$L_{tr}=0.0$\%&$L_{tr}=0.0$\%\\
&&$N_{sv}$=139&&$N_{sv}$=1084&$N_{sv}$=178&$N_{sv}$=14&$N_{sv}$=13&$N_{sv}$=11&$N_{sv}$=9\\\hline
Krapivsky &5/7&$f(\bx)=-2.32$&$f(\bx)=-2.44$&&$f(\bx)=2.56$&$f(\bx)=0.95$&$f(\bx)=2.67$&$f(\bx)=2.69$&$f(\bx)=1.72$\\
&&$L_{tst}=4.5$\%&$L_{tst}=32.8$\%&&$L_{tst}=0.8$\%&$L_{tst}=0.0$\%&$L_{tst}=0.0$\%&$L_{tst}=0.0$\%&$L_{tst}=0.0$\%\\
&&$L_{tr}=3.2$\%&$L_{tr}=31.3$\%&&$L_{tr}=1.6$\%&$L_{tr}=0.0$\%&$L_{tr}=0.0$\%&$L_{tr}=0.0$\%&$L_{tr}=0.0$\%\\
&&$N_{sv}$=122&$N_{sv}$=1084&&$N_{sv}$=223&$N_{sv}$=12&$N_{sv}$=13&$N_{sv}$=12&$N_{sv}$=9\\\hline
Kim &4/7&$f(\bx)=-2.80$&$f(\bx)=-2.49$&$f(\bx)=-2.56$&&$f(\bx)=0.36$&$f(\bx)=0.87$&$f(\bx)=1.53$&$f(\bx)=1.06$\\
&&$L_{tst}=0.8$\%&$L_{tst}=0.8$\%&$L_{tst}=0.8$\%&&$L_{tst}=0.0$\%&$L_{tst}=9.0$\%&$L_{tst}=3.0$\%&$L_{tst}=0.0$\%\\
&&$L_{tr}=0.7$\%&$L_{tr}=0.9$\%&$L_{tr}=1.6$\%&&$L_{tr}=0.0$\%&$L_{tr}=10.5$\%&$L_{tr}=2.9$\%&$L_{tr}=0.1$\%\\
&&$N_{sv}$=194&$N_{sv}$=178&$N_{sv}$=223&&$N_{sv}$=47&$N_{sv}$=498&$N_{sv}$=180&$N_{sv}$=84\\\hline
Vazquez &3/7&$f(\bx)=-1.12$&$f(\bx)=-1.01$&$f(\bx)=-0.95$&$f(\bx)=-0.36$&&$f(\bx)=0.60$&$f(\bx)=1.25$&$f(\bx)=1.23$\\
&&$L_{tst}=0.0$\%&$L_{tst}=0.0$\%&$L_{tst}=0.0$\%&$L_{tst}=0.0$\%&&$L_{tst}=0.0$\%&$L_{tst}=0.0$\%&$L_{tst}=0.0$\%\\
&&$L_{tr}=0.0$\%&$L_{tr}=0.0$\%&$L_{tr}=0.0$\%&$L_{tr}=0.0$\%&&$L_{tr}=0.0$\%&$L_{tr}=0.0$\%&$L_{tr}=0.0$\%\\
&&$N_{sv}$=9&$N_{sv}$=14&$N_{sv}$=12&$N_{sv}$=47&&$N_{sv}$=8&$N_{sv}$=6&$N_{sv}$=10\\\hline
Erdos &2/7&$f(\bx)=-3.58$&$f(\bx)=-2.33$&$f(\bx)=-2.67$&$f(\bx)=-0.87$&$f(\bx)=-0.60$&&$f(\bx)=1.43$&$f(\bx)=1.36$\\
&&$L_{tst}=0.0$\%&$L_{tst}=0.0$\%&$L_{tst}=0.0$\%&$L_{tst}=9.0$\%&$L_{tst}=0.0$\%&&$L_{tst}=2.3$\%&$L_{tst}=0.0$\%\\
&&$L_{tr}=0.0$\%&$L_{tr}=0.0$\%&$L_{tr}=0.0$\%&$L_{tr}=10.5$\%&$L_{tr}=0.0$\%&&$L_{tr}=2.3$\%&$L_{tr}=0.0$\%\\
&&$N_{sv}$=10&$N_{sv}$=13&$N_{sv}$=13&$N_{sv}$=498&$N_{sv}$=8&&$N_{sv}$=130&$N_{sv}$=7\\\hline
Grindrod &1/7&$f(\bx)=-3.11$&$f(\bx)=-2.30$&$f(\bx)=-2.69$&$f(\bx)=-1.53$&$f(\bx)=-1.25$&$f(\bx)=-1.43$&&$f(\bx)=1.37$\\
&&$L_{tst}=0.0$\%&$L_{tst}=0.0$\%&$L_{tst}=0.0$\%&$L_{tst}=3.0$\%&$L_{tst}=0.0$\%&$L_{tst}=2.3$\%&&$L_{tst}=0.0$\%\\
&&$L_{tr}=0.0$\%&$L_{tr}=0.0$\%&$L_{tr}=0.0$\%&$L_{tr}=2.9$\%&$L_{tr}=0.0$\%&$L_{tr}=2.3$\%&&$L_{tr}=0.0$\%\\
&&$N_{sv}$=9&$N_{sv}$=11&$N_{sv}$=12&$N_{sv}$=180&$N_{sv}$=6&$N_{sv}$=130&&$N_{sv}$=12\\\hline
MZ &0/7&$f(\bx)=-1.26$&$f(\bx)=-1.64$&$f(\bx)=-1.72$&$f(\bx)=-1.06$&$f(\bx)=-1.23$&$f(\bx)=-1.36$&$f(\bx)=-1.37$&\\
&&$L_{tst}=0.0$\%&$L_{tst}=0.0$\%&$L_{tst}=0.0$\%&$L_{tst}=0.0$\%&$L_{tst}=0.0$\%&$L_{tst}=0.0$\%&$L_{tst}=0.0$\%&\\
&&$L_{tr}=0.0$\%&$L_{tr}=0.0$\%&$L_{tr}=0.0$\%&$L_{tr}=0.1$\%&$L_{tr}=0.0$\%&$L_{tr}=0.0$\%&$L_{tr}=0.0$\%&\\
&&$N_{sv}$=9&$N_{sv}$=9&$N_{sv}$=9&$N_{sv}$=84&$N_{sv}$=10&$N_{sv}$=7&$N_{sv}$=12&\\\hline
\end{tabular}
}
\end{center}\caption{SVM results for E. coli. $f(x)={\bf w}\cdot \bx_{E. coli} +b$, $L_{tst}$ is the test loss, $L_{tr}$
the training loss and $N_{sv}$ the number of support vectors.
Results are shown for SVMs trained between every pair of models.
if $f(x)>0$ E. coli is classified as the row-header, if $f(x)<0$
as the column-header.} \label{svmcoli}\end{sidewaystable}

\begin{sidewaystable}[h]\begin{center}
\small{
\begin{tabular}{|p{.8 in}|l|l|l|l|l|l|l|l|}
\hline
&Kumar&Krapivsky-Bianconi&Krapivksy&Kim&Vazquez&Erdos&Grindrod&MZ\\\hline
Kumar &&sum(ATA)&sum(ATA)&nnz(ATA)&nnz D(AATA)&nnz(ATA)&nnz(ATA)&nnz(AA)\\
&&$L_{tst}=0.3$\%&$L_{tst}=0.0$\%&$L_{tst}=0.0$\%&$L_{tst}=0.0$\%&$L_{tst}=0.0$\%&$L_{tst}=0.0$\%&$L_{tst}=0.0$\%\\
&&$L_{tr}=0.1$\%&$L_{tr}=0.4$\%&$L_{tr}=0.0$\%&$L_{tr}=0.0$\%&$L_{tr}=0.0$\%&$L_{tr}=0.0$\%&$L_{tr}=0.0$\%\\\hline
Krapivsky-Bianconi &&&nnz(ADATA)&nnz(ATA)&nnz D(ATA)&nnz(ATA)&nnz(ATA)&sum D(AA)\\
&&&$L_{tst}=27.8$\%&$L_{tst}=0.0$\%&$L_{tst}=0.0$\%&$L_{tst}=0.0$\%&$L_{tst}=0.0$\%&$L_{tst}=0.0$\%\\
&&&$L_{tr}=26.9$\%&$L_{tr}=0.0$\%&$L_{tr}=0.0$\%&$L_{tr}=0.0$\%&$L_{tr}=0.0$\%&$L_{tr}=0.1$\%\\\hline
Krapivksy &&&&nnz(ATA)&nnz D(AATA)&nnz(ATA)&nnz(ATA)&sum D(AA)\\
&&&&$L_{tst}=0.0$\%&$L_{tst}=0.0$\%&$L_{tst}=0.0$\%&$L_{tst}=0.0$\%&$L_{tst}=0.0$\%\\
&&&&$L_{tr}=0.0$\%&$L_{tr}=0.0$\%&$L_{tr}=0.0$\%&$L_{tr}=0.0$\%&$L_{tr}=0.0$\%\\\hline
Kim &&&&&nnz(ATA)&sum U(AAUATA)&sum U(AUTAA)&sum D(AADAA)\\
&&&&&$L_{tst}=0.0$\%&$L_{tst}=7.8$\%&$L_{tst}=5.0$\%&$L_{tst}=4.0$\%\\
&&&&&$L_{tr}=0.0$\%&$L_{tr}=10.1$\%&$L_{tr}=5.6$\%&$L_{tr}=4.6$\%\\\hline
Vazquez &&&&&&nnz(ATA)&nnz(ATA)&nnz(AA)\\
&&&&&&$L_{tst}=0.0$\%&$L_{tst}=0.0$\%&$L_{tst}=0.0$\%\\
&&&&&&$L_{tr}=0.0$\%&$L_{tr}=0.0$\%&$L_{tr}=0.0$\%\\\hline
Erdos &&&&&&&sum D(AADATA)&nnz(AA)\\
&&&&&&&$L_{tst}=2.5$\%&$L_{tst}=0.0$\%\\
&&&&&&&$L_{tr}=2.8$\%&$L_{tr}=0.0$\%\\\hline
Grindrod &&&&&&&&nnz(AA)\\
&&&&&&&&$L_{tst}=0.0$\%\\
&&&&&&&&$L_{tr}=0.0$\%\\\hline
MZ &&&&&&&&\\
&&&&&&&&\\
&&&&&&&&\\\hline
\end{tabular}
}
\end{center}\caption{Most discriminative words for the E. coli training data based on lowest test loss by 1-dimensional splitting
for every pair of models. $L_{tst}$ is the test loss and $L_{tr}$
the training loss.}\label{ecolisingle}\end{sidewaystable}

\begin{table}[h]\begin{center}
\small{
\begin{tabular}{|c|c|c|c|}
\hline RANKING&WORD& $L_{tr}$&ASSOCIATED SUBGRAPHS
\\\hline 1&sum U(ATA)&5.8\% & \includegraphics[height = 20  mm ]{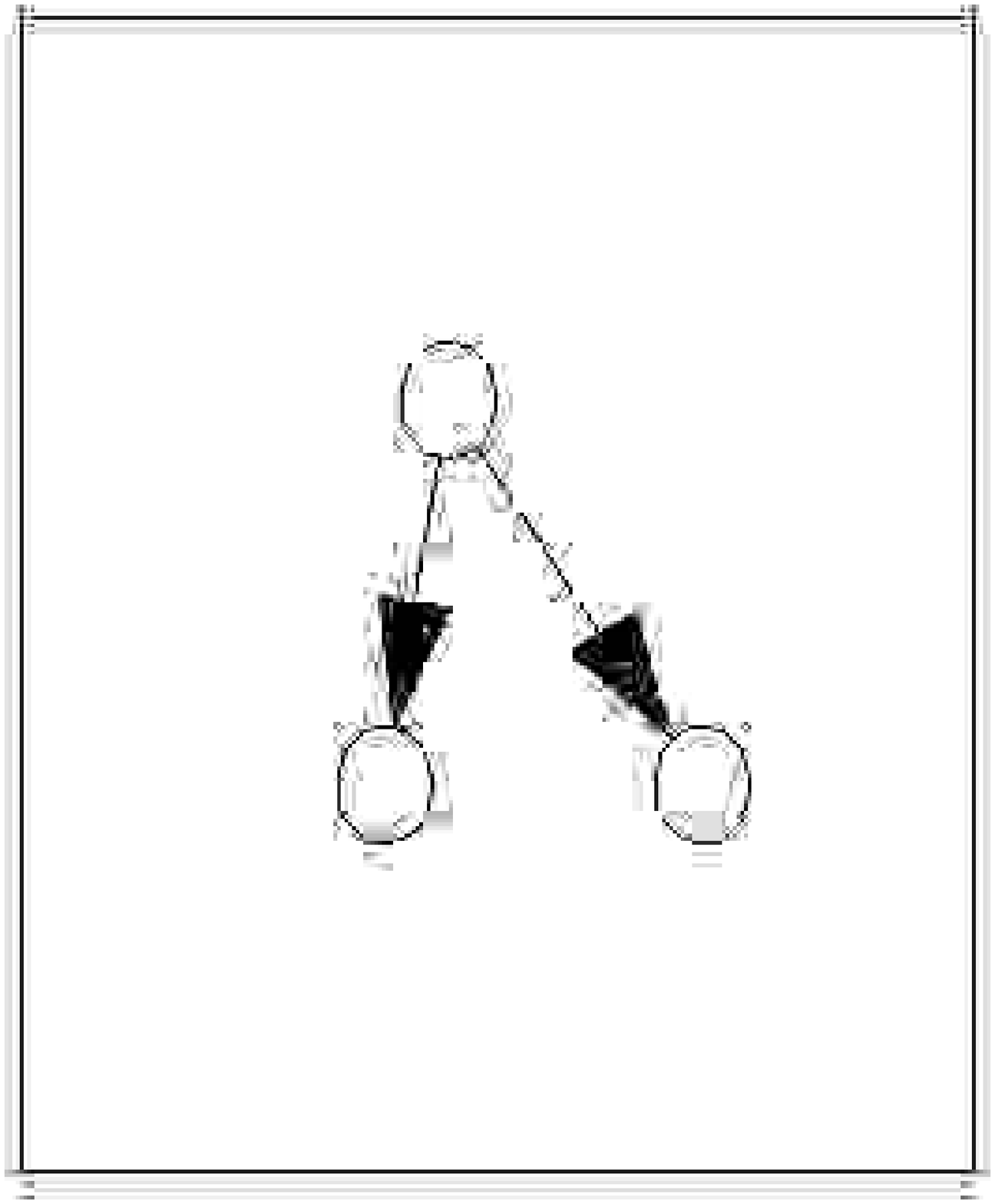}
\\\hline2&nnz D(AUTA)&2.4\%& \includegraphics[height = 20  mm ]{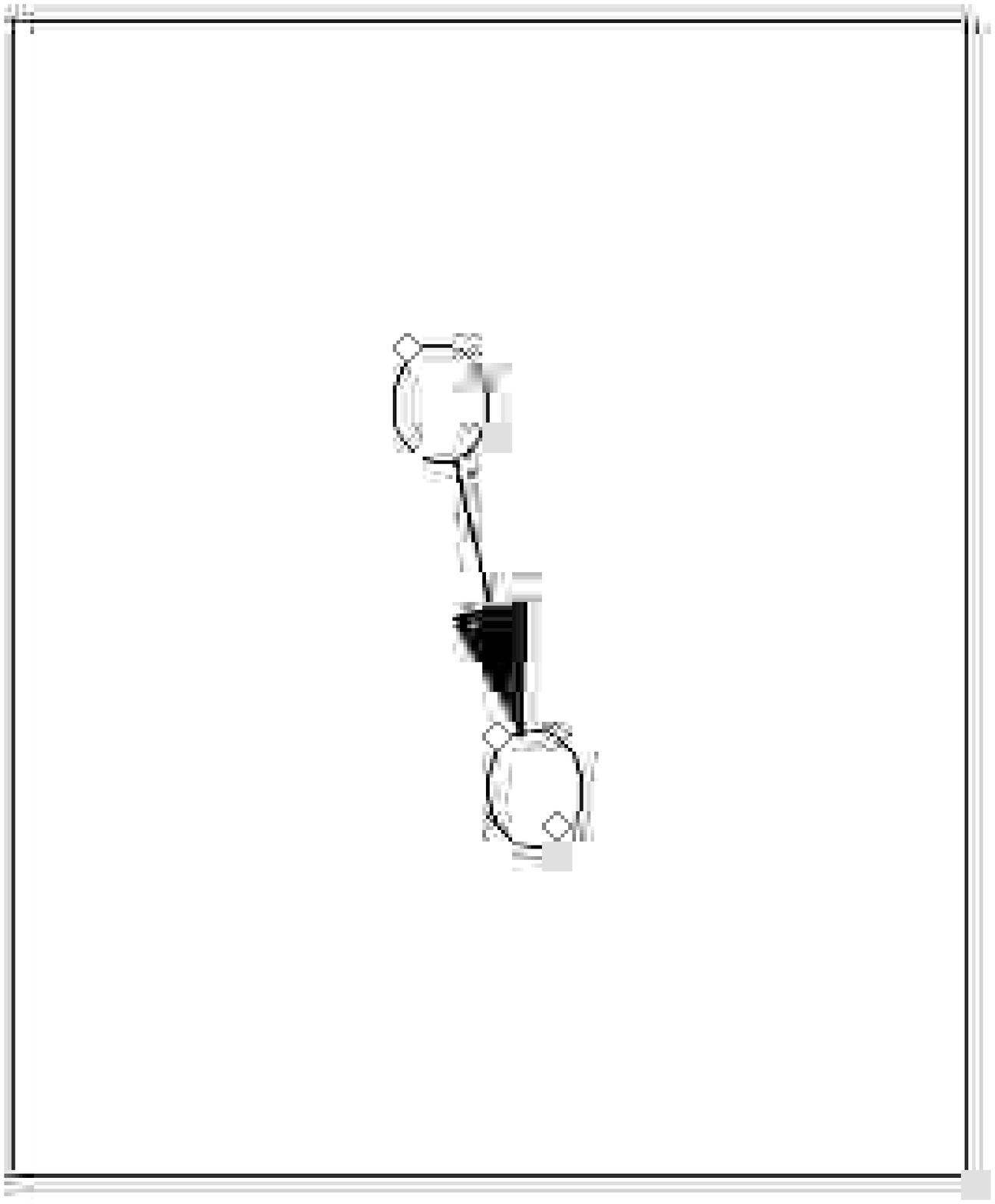}
\\\hline 3&sum D(AADAA)&1.7\%& \includegraphics[height = 20  mm ]{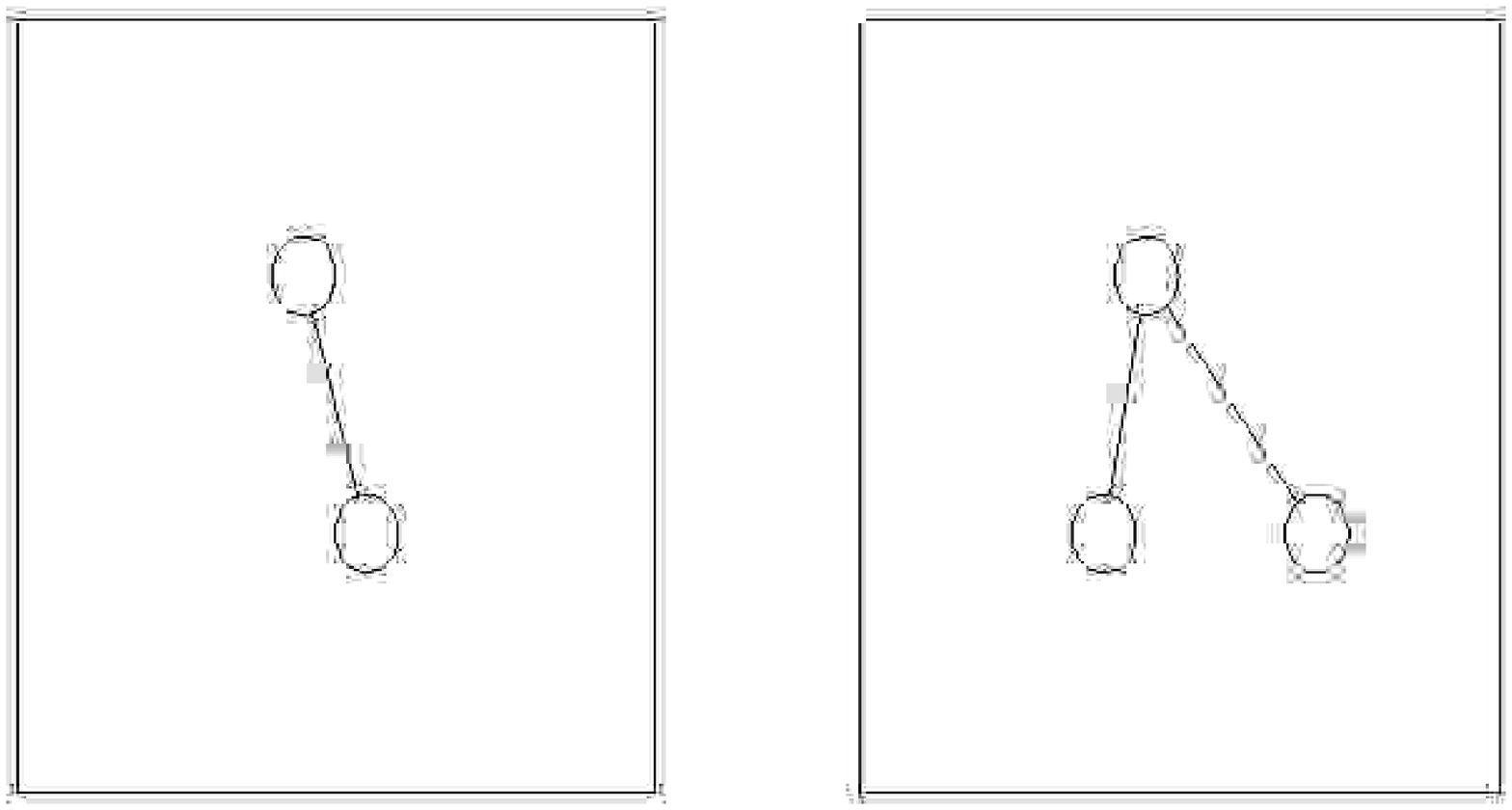}
\\\hline 4&nnzU(AUAUTAUTA)&1.4\%&\includegraphics[height = 20  mm ]{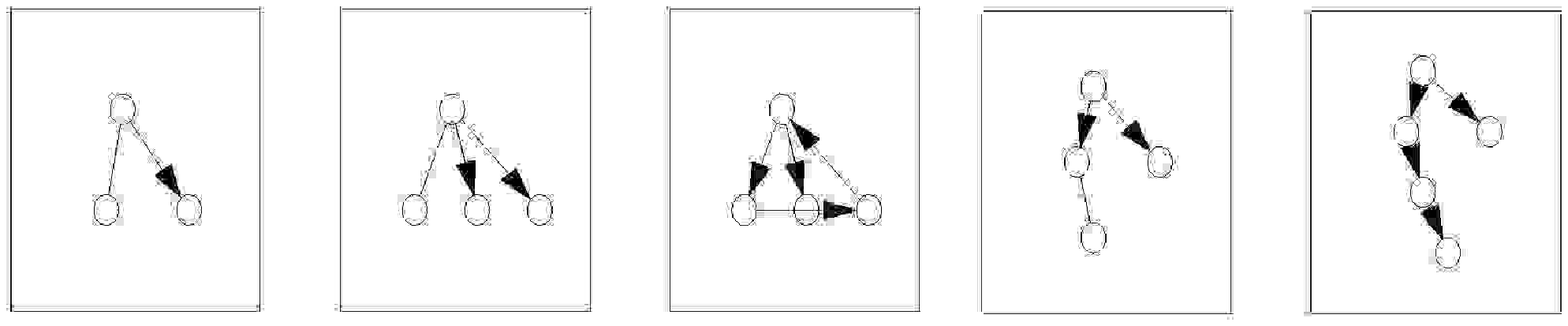}
\\\hline 5&nnz D(AUTAUAUTAUA)&1.3\%&\includegraphics[height = 20  mm ]{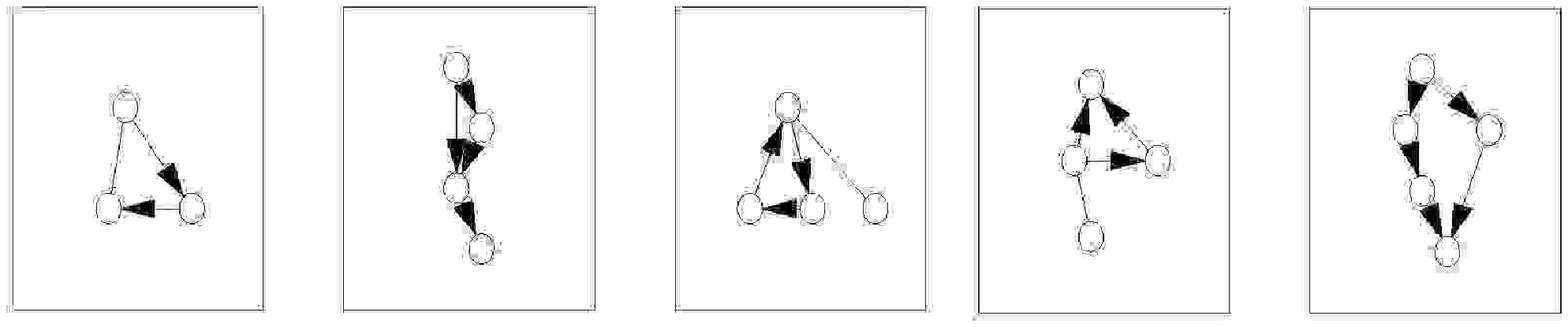}
\\\hline6&nnz U(ADAUTA)&1.2\%&\includegraphics[height = 20  mm ]{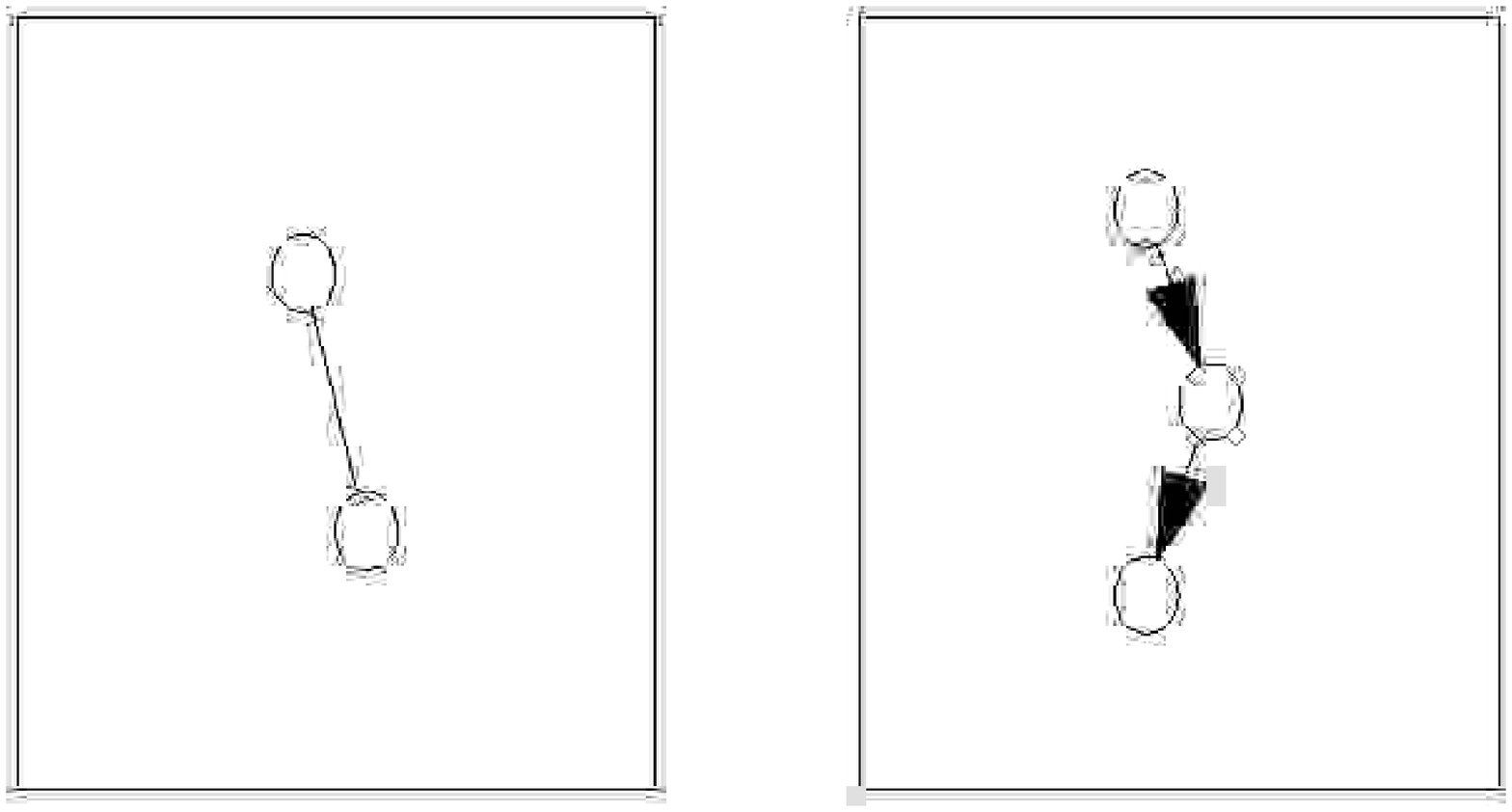}
\\\hline 7&sum D(AUTAUTAUTA)&1.2\%&\includegraphics[height = 20  mm ]{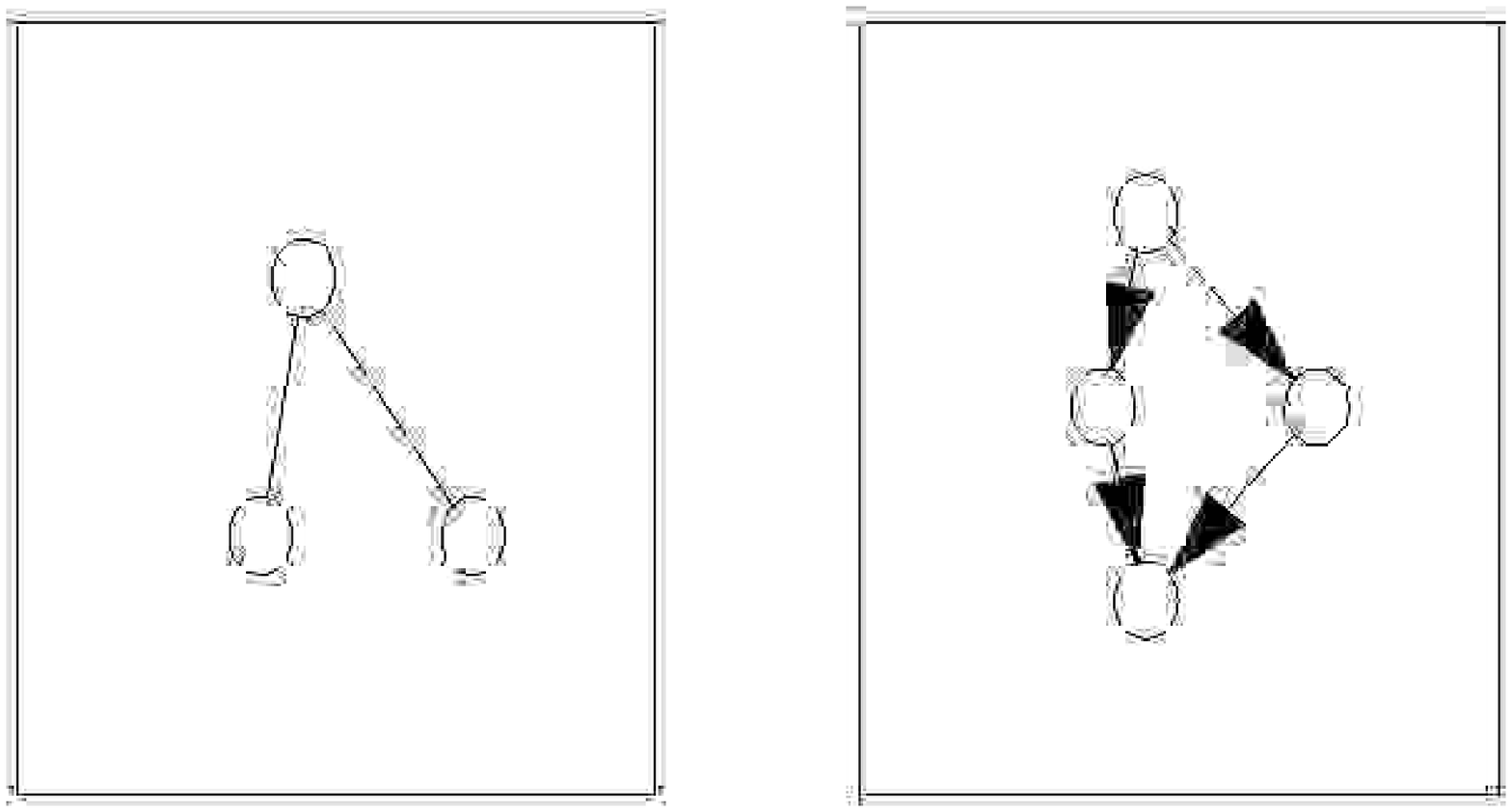}
\\\hline8&nnz U(AAUA)&1.1\%&\includegraphics[height = 20  mm ]{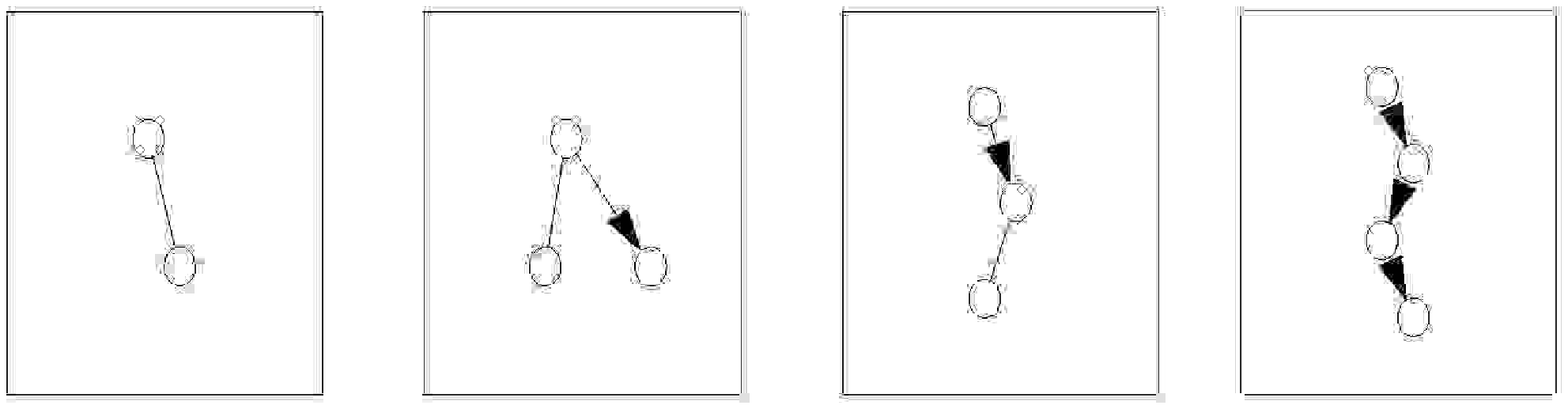}
\\\hline 9&sum(AUTA)&1.1\%&\includegraphics[height = 20  mm ]{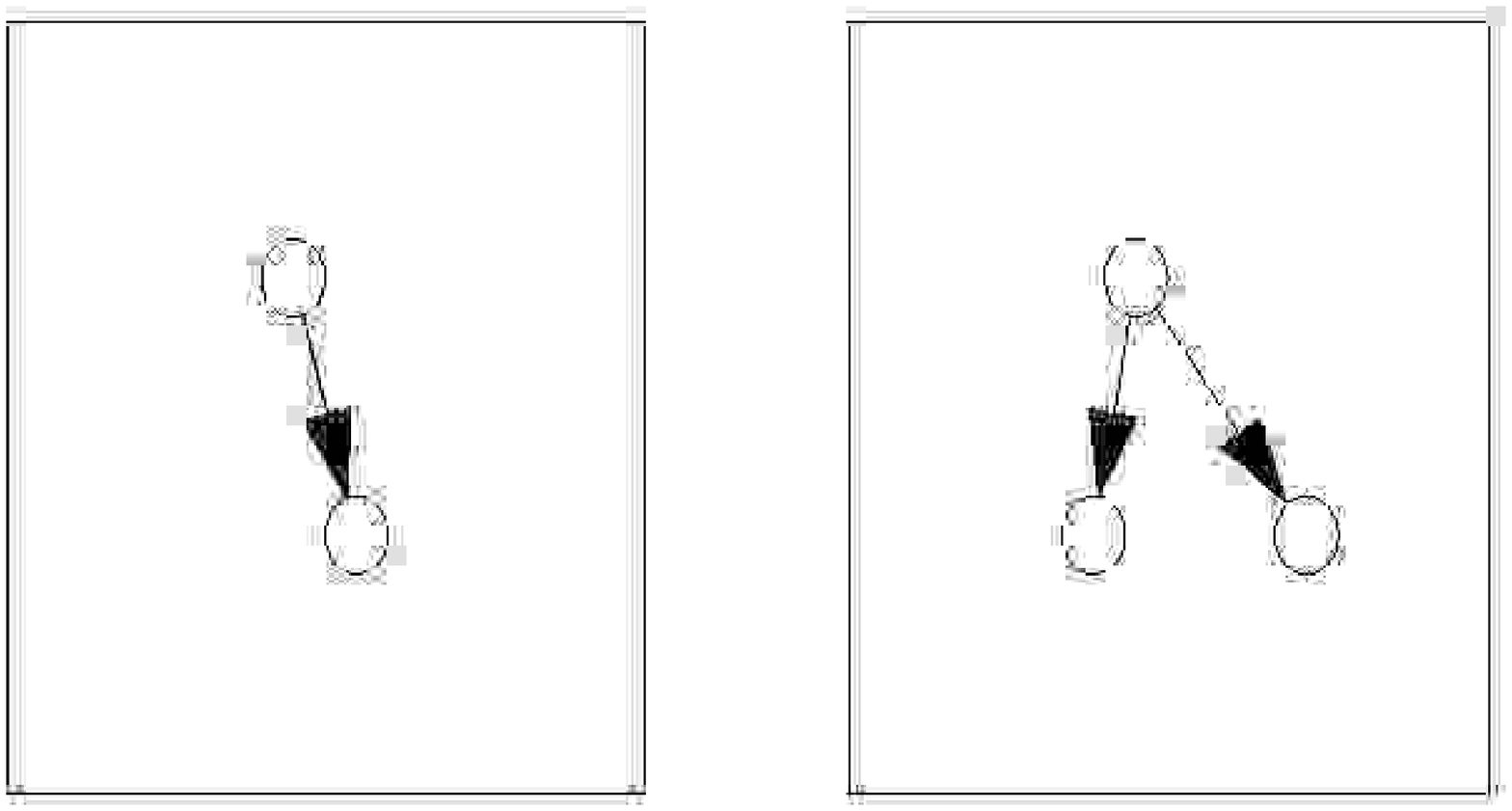}
\\\hline 10&sumU(ADAUADAUTA)&1.0\%&\includegraphics[height = 20  mm ]{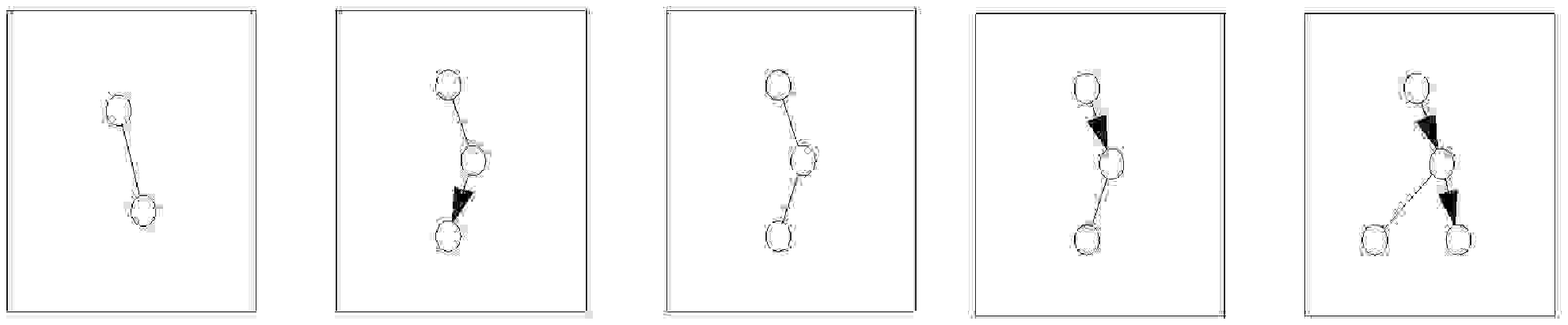}
\\\hline

\end{tabular}
}
\end{center}\caption{
Ranking of words found by binary pairwise trees for the E. coli
training data. $L_{tr}$ for a word ranked $n$ is the average
training loss over all pairwise trees, where every tree has depth
$n$ and splits the data using words $1$ to $n$ in the given order.
%See Sec. \ref{sec:wordranking} for details.
}\label{colimytree}\end{table}

\begin{sidewaystable}[h]\begin{center}
\small{
\begin{tabular}{|l|l|l|l|l|l|l|l|l|}
\hline
&MZ&Grindrod&Kim&Erdos&Kumar&Krapivsky-Bianconi&Vazquez&Krapivksy\\\hline
MZ &&sum(AA)&nnz D(AAAAUA)&sum(AA)&nnz(AA)&sum D(AATA)&nnz(AA)&sum D(AATA)\\
&&$L_{tst}=0.0$\%&$L_{tst}=3.8$\%&$L_{tst}=0.0$\%&$L_{tst}=0.0$\%&$L_{tst}=0.0$\%&$L_{tst}=0.0$\%&$L_{tst}=0.0$\%\\
&&$L_{tr}=0.0$\%&$L_{tr}=4.3$\%&$L_{tr}=0.0$\%&$L_{tr}=0.0$\%&$L_{tr}=0.0$\%&$L_{tr}=0.0$\%&$L_{tr}=0.0$\%\\\hline
Grindrod &&&sum(ATADATA)&sum D(AAUATA)&nnz D(AA)&sum(AA)&nnz(AA)&sum(AA)\\
&&&$L_{tst}=3.8$\%&$L_{tst}=1.5$\%&$L_{tst}=0.0$\%&$L_{tst}=0.0$\%&$L_{tst}=0.0$\%&$L_{tst}=0.0$\%\\
&&&$L_{tr}=5.1$\%&$L_{tr}=1.3$\%&$L_{tr}=0.0$\%&$L_{tr}=0.0$\%&$L_{tr}=0.0$\%&$L_{tr}=0.0$\%\\\hline
Kim &&&&sum D(ATAUATA)&nnz D(AA)&nnz D(ATA)&nnz(AA)&nnz D(ATA)\\
&&&&$L_{tst}=1.0$\%&$L_{tst}=0.0$\%&$L_{tst}=0.0$\%&$L_{tst}=0.0$\%&$L_{tst}=0.0$\%\\
&&&&$L_{tr}=2.3$\%&$L_{tr}=0.0$\%&$L_{tr}=0.0$\%&$L_{tr}=0.0$\%&$L_{tr}=0.0$\%\\\hline
Erdos &&&&&nnz(AA)&sum(AA)&nnz(AA)&sum(AA)\\
&&&&&$L_{tst}=0.0$\%&$L_{tst}=0.0$\%&$L_{tst}=0.0$\%&$L_{tst}=0.0$\%\\
&&&&&$L_{tr}=0.0$\%&$L_{tr}=0.0$\%&$L_{tr}=0.0$\%&$L_{tr}=0.0$\%\\\hline
Kumar &&&&&&nnz D(AA)&nnz(AA)&nnz D(AA)\\
&&&&&&$L_{tst}=0.0$\%&$L_{tst}=0.0$\%&$L_{tst}=0.0$\%\\
&&&&&&$L_{tr}=0.0$\%&$L_{tr}=0.0$\%&$L_{tr}=0.0$\%\\\hline
Krapivsky-Bianconi &&&&&&&nnz(AA)&nnz(AA)\\
&&&&&&&$L_{tst}=0.0$\%&$L_{tst}=16.5$\%\\
&&&&&&&$L_{tr}=0.0$\%&$L_{tr}=15.4$\%\\\hline
Vazquez &&&&&&&&nnz(AA)\\
&&&&&&&&$L_{tst}=0.0$\%\\
&&&&&&&&$L_{tr}=0.0$\%\\\hline
Krapivksy &&&&&&&&\\
&&&&&&&&\\
&&&&&&&&\\\hline
\end{tabular}
}
\end{center}\caption{
Most discriminative words for the C. elegans training data based
on lowest test loss by 1-dimensional splitting for every pair of
models. $L_{tst}$ is the test loss and $L_{tr}$ the training
loss.} \label{celegsingle}
\end{sidewaystable}

\begin{table}[h]\begin{center}\begin{tabular}{|c|c|c|c|}
\hline RANKING&WORD& $L_{tr}$&ASSOCIATED SUBGRAPHS
\\\hline 1&sumD(AAUAUAA)&3.6\%& \includegraphics[height = 20  mm ]{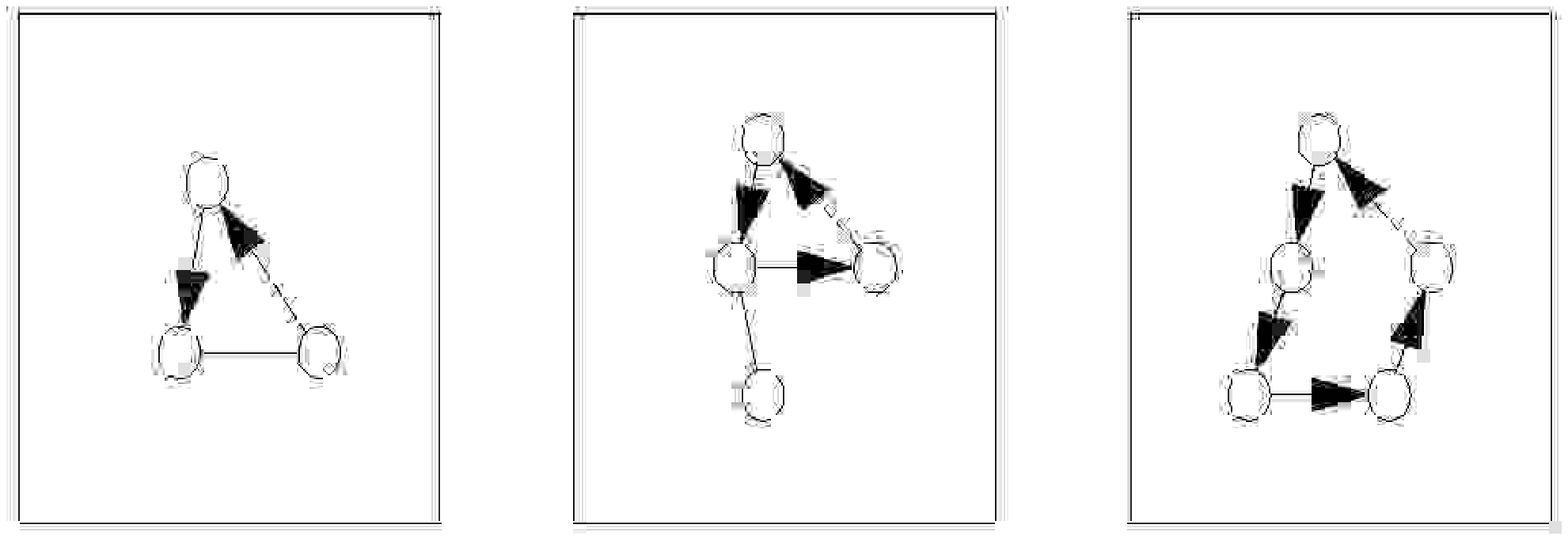}
\\\hline 2&nnz D(AUAUA)&1.1\%&\includegraphics[height = 20  mm ]{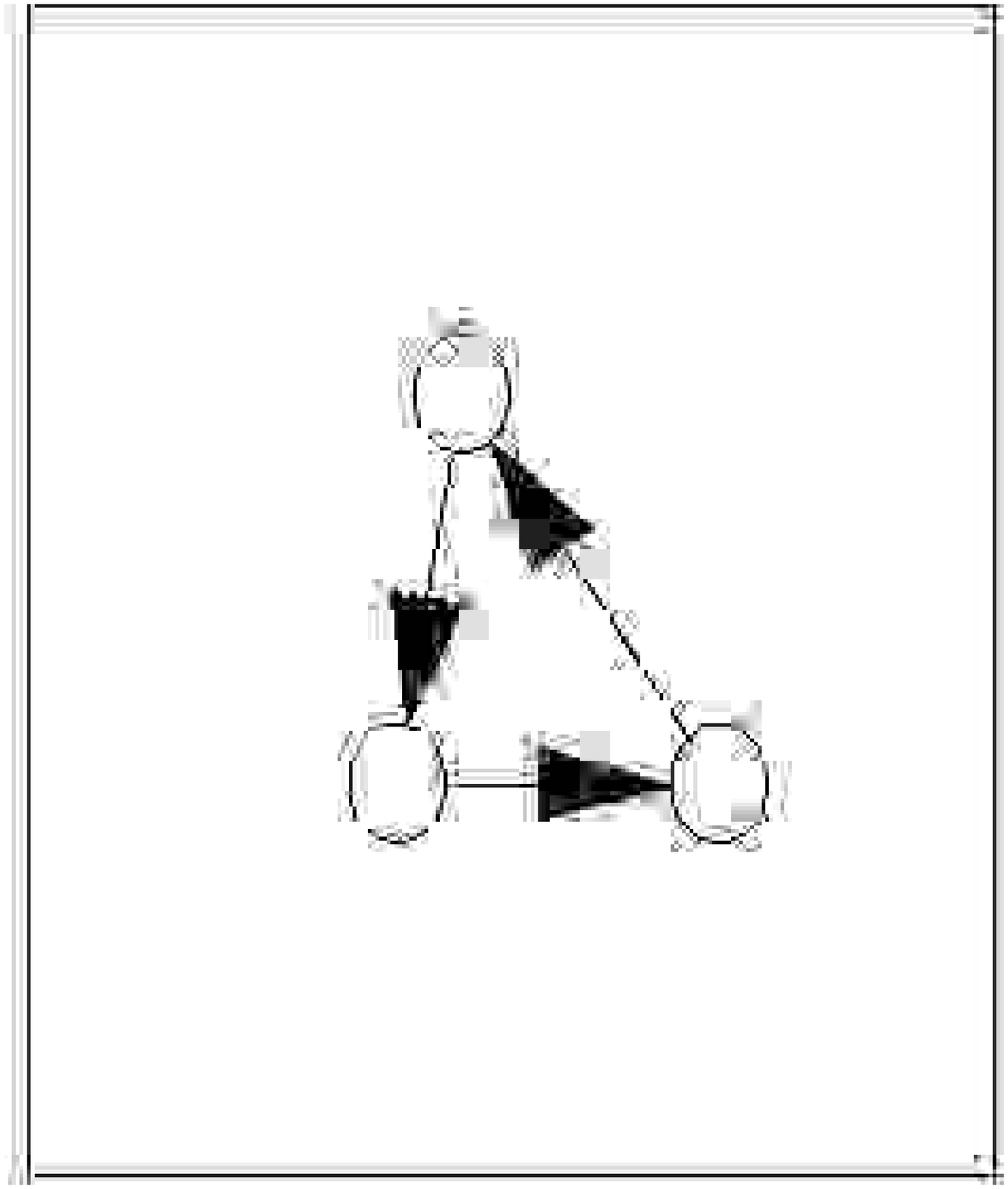}
\\\hline 3&sumU(AUTA)&0.8\%&\includegraphics[height = 20  mm ]{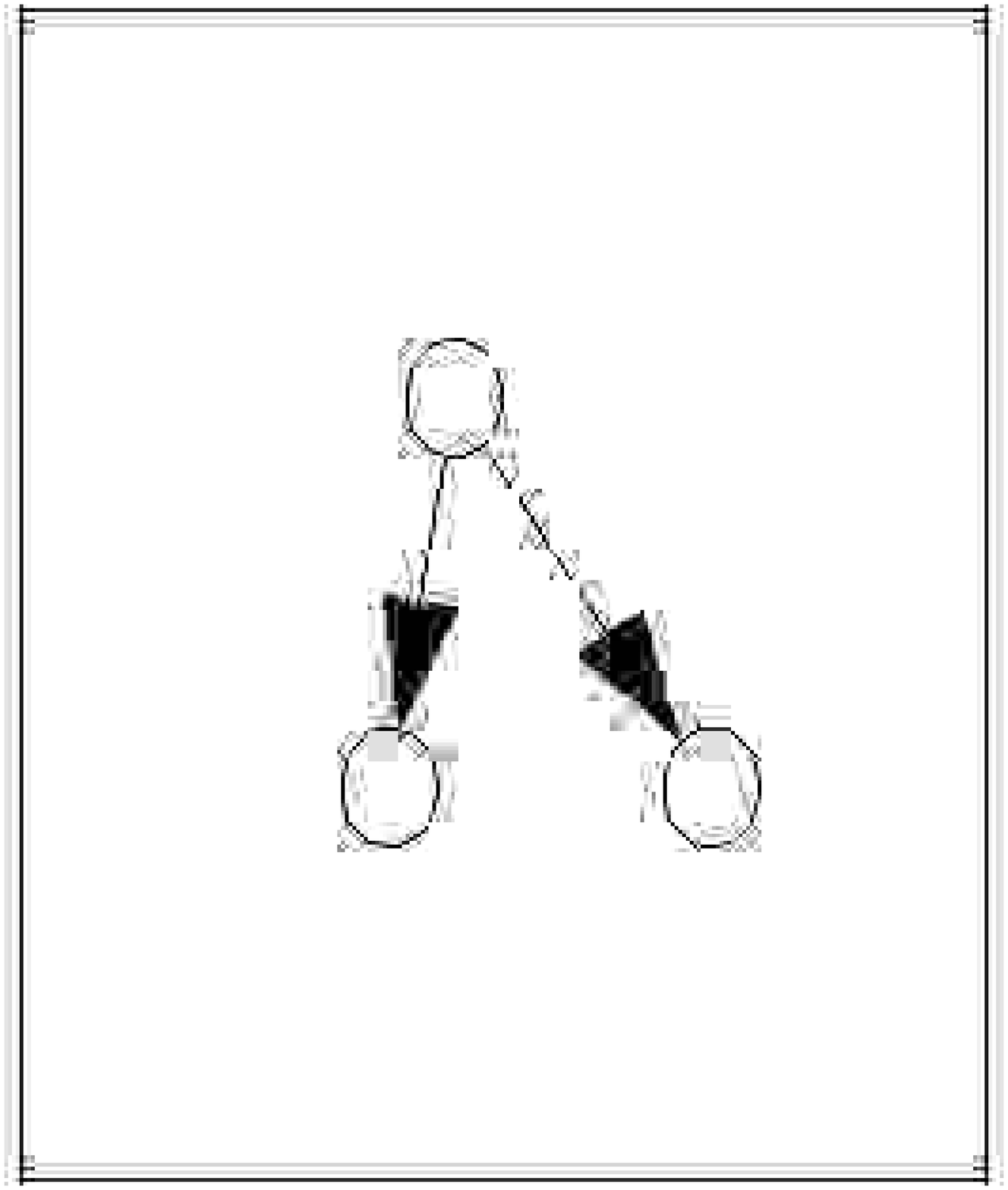}
\\\hline 4&sum D(AUAUTAUAUTA)&0.6\%&\includegraphics[height = 20  mm ]{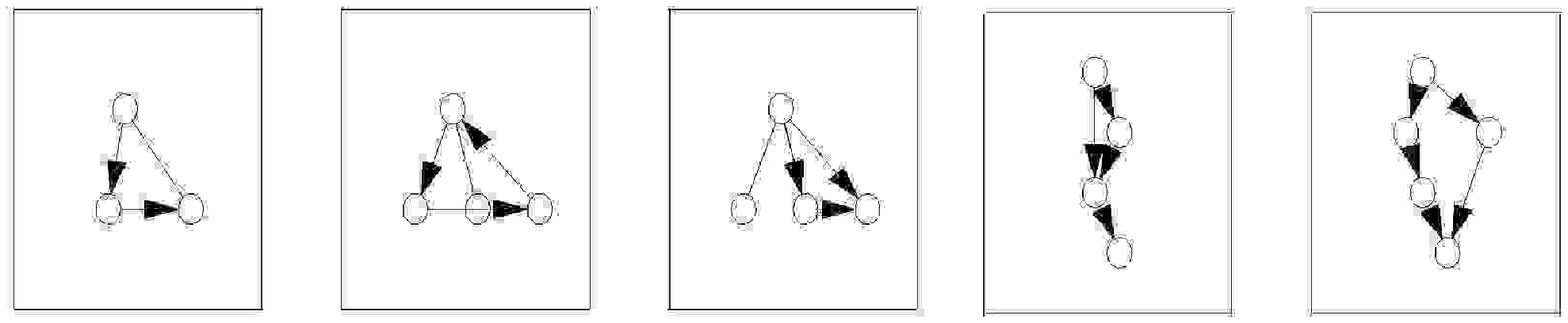}
\\\hline 5&nnzU(AUA)&0.5\%&\includegraphics[height = 20  mm ]{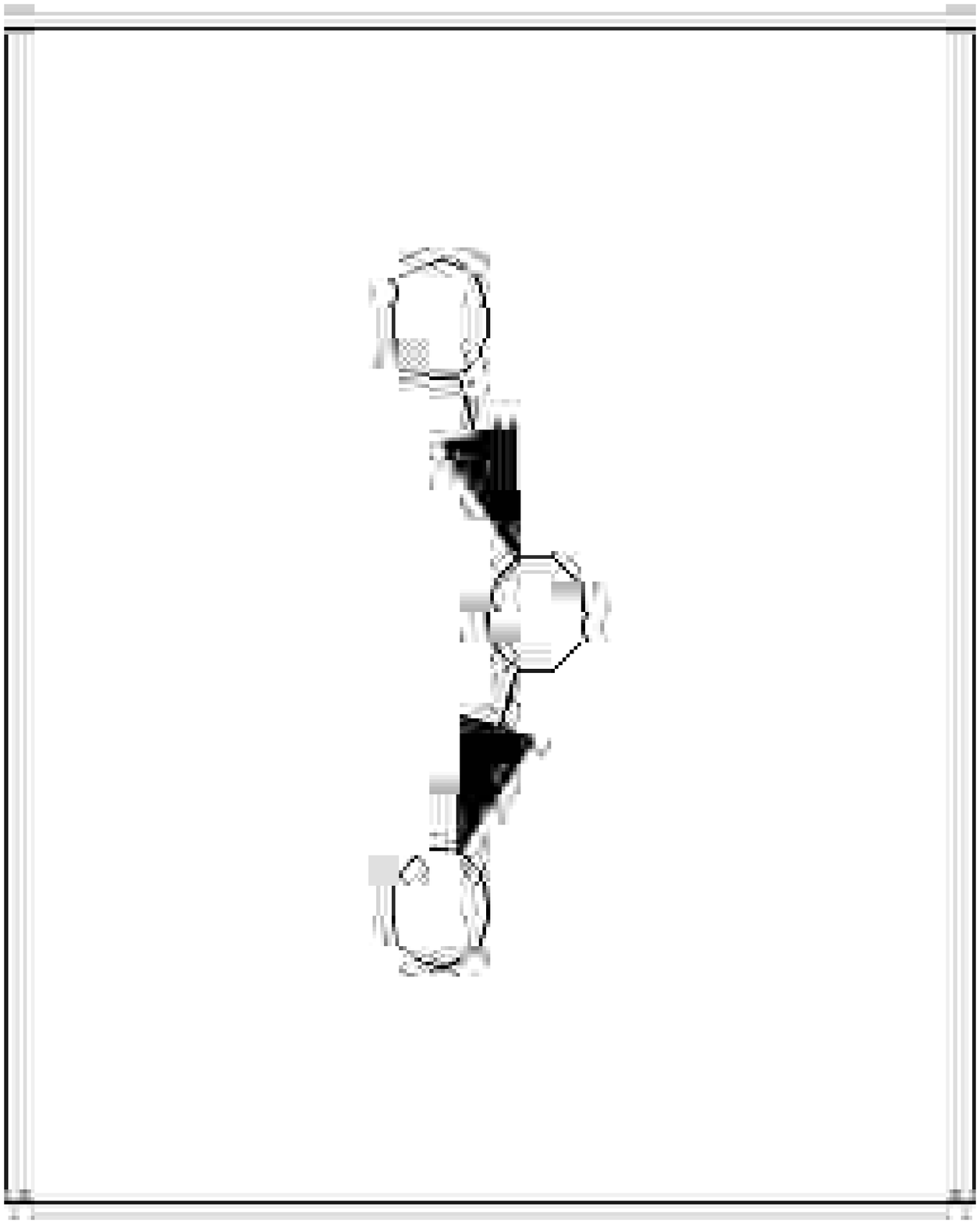}
\\\hline 6&sum D(AA)&0.5\%&\includegraphics[height = 20  mm ]{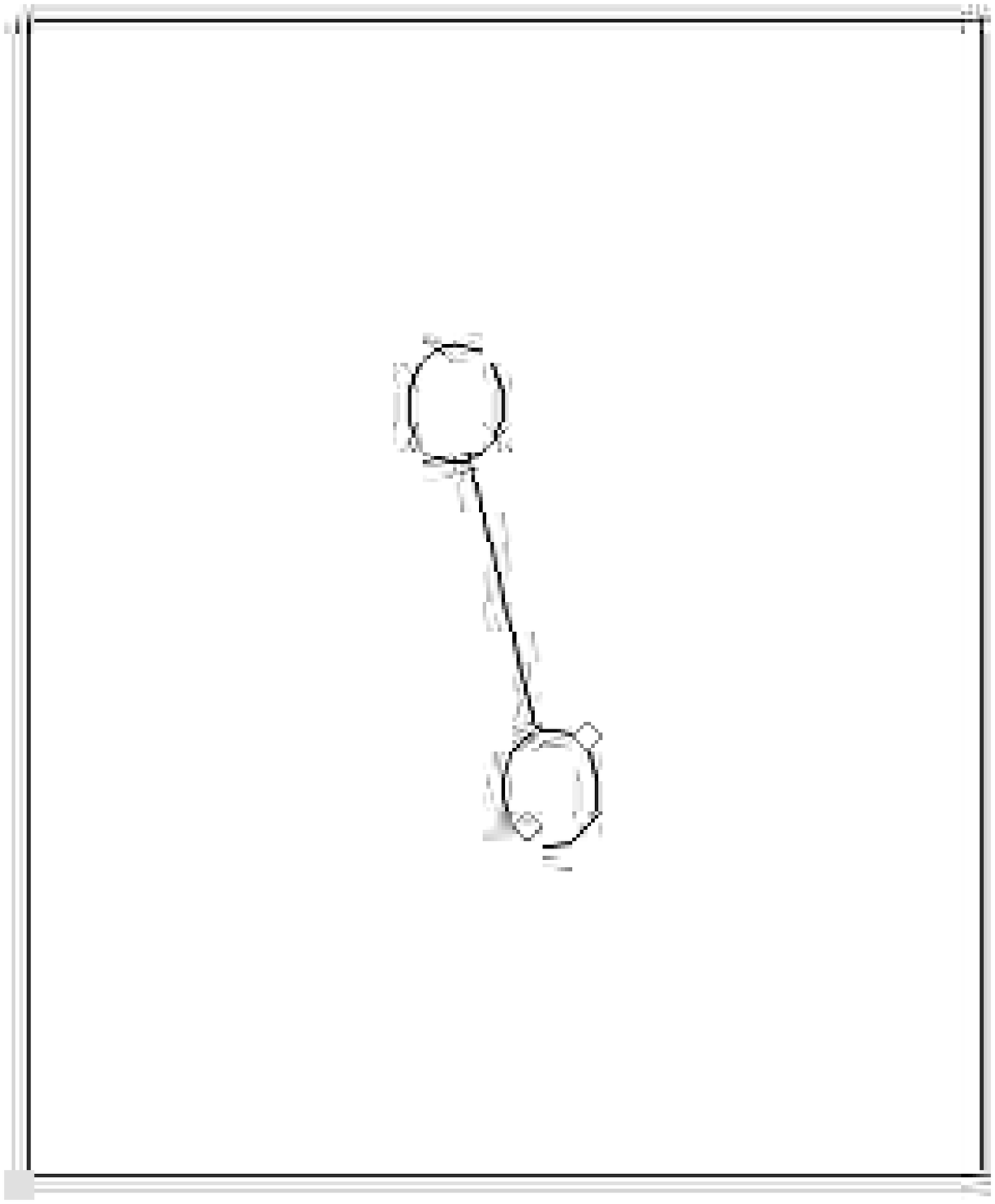}
\\\hline 7&sumU(AUTADAUAUA)&0.4\%&\includegraphics[height = 20  mm ]{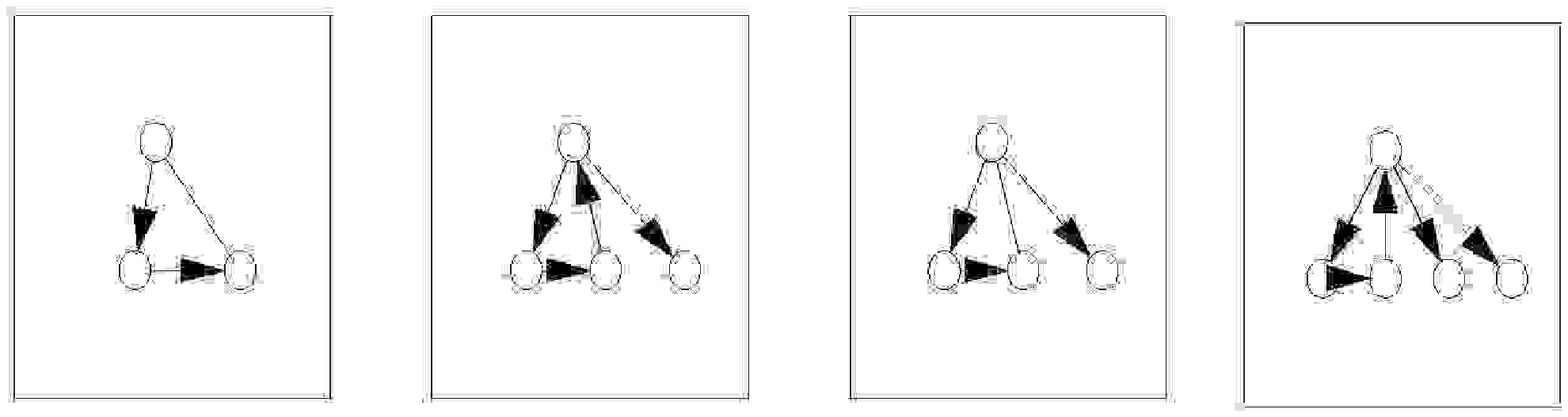}
\\\hline 8&nnz U(ADAUTA)&0.4\%&\includegraphics[height = 20  mm ]{nnzUADAUTA}
\\\hline 9&nnzD(AUADATAUA)&0.4\%&\includegraphics[height = 20  mm ]{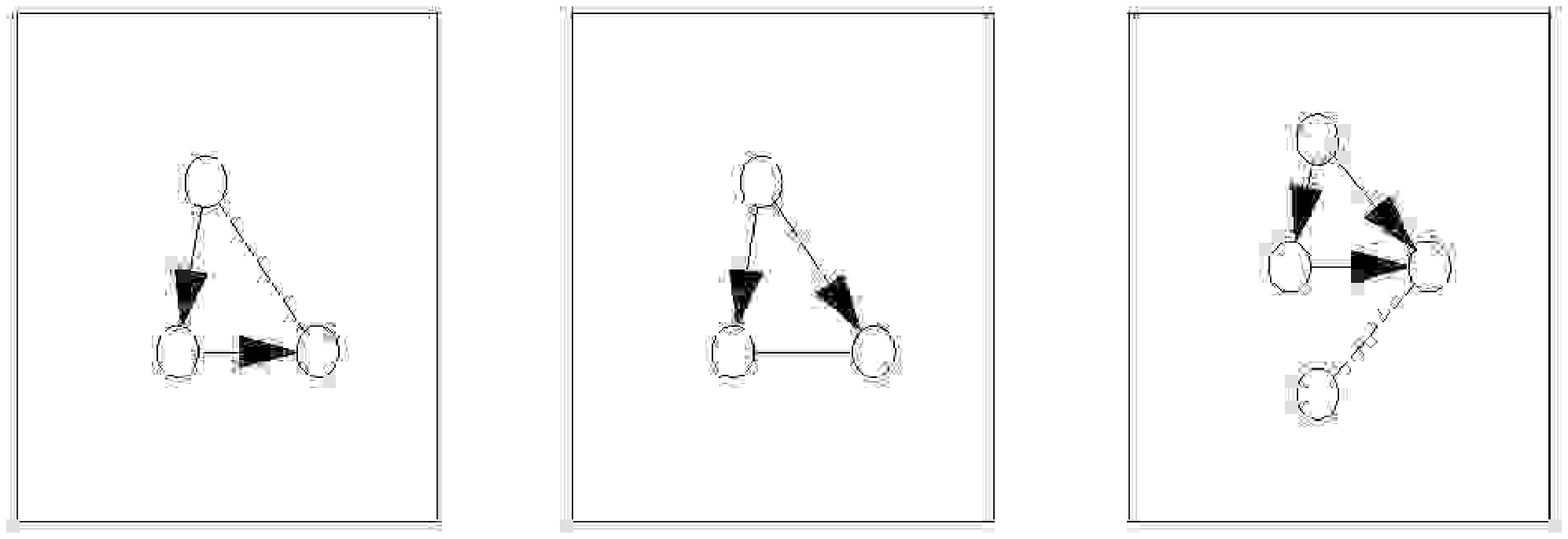}
\\\hline
\end{tabular}\end{center}\caption{
Ranking of words found by binary pairwise trees for the C. elegans
training data. $L_{tr}$ for a word ranked $n$ is the average
training loss over all pairwise trees, where every tree has depth
$n$ and splits the data using words $1$ to $n$ in the given order.
%See Sec. \ref{sec:wordranking} for details.
}\label{celegmytree}\end{table}

\begin{sidewaystable}[h]\begin{center}
\footnotesize{
\begin{tabular}
{|l|l|l|l|l|l|l|l|l|l|}
\hline
&votes&MZ&Kim&Grindrod&Krapivsky-Bianconi&Krapivksy&Erdos&Kumar&Vazquez\_K5\\\hline
MZ &7/7&&$f(\bx)=1.82$&$f(\bx)=0.49$&$f(\bx)=3.19$&$f(\bx)=2.28$&$f(\bx)=1.18$&$f(\bx)=0.91$&$f(\bx)=1.25$\\
&&&$L_{tst}=0.0$\%&$L_{tst}=0.0$\%&$L_{tst}=0.0$\%&$L_{tst}=0.0$\%&$L_{tst}=0.0$\%&$L_{tst}=0.0$\%&$L_{tst}=0.0$\%\\
&&&$L_{tr}=0.0$\%&$L_{tr}=0.0$\%&$L_{tr}=0.0$\%&$L_{tr}=0.0$\%&$L_{tr}=0.0$\%&$L_{tr}=0.0$\%&$L_{tr}=0.0$\%\\
&&&$N_{sv}$=48&$N_{sv}$=11&$N_{sv}$=48&$N_{sv}$=37&$N_{sv}$=10&$N_{sv}$=8&$N_{sv}$=5\\\hline
Kim &6/7&$f(\bx)=-1.82$&&$f(\bx)=0.99$&$f(\bx)=0.63$&$f(\bx)=0.06$&$f(\bx)=16.99$&$f(\bx)=1.25$&$f(\bx)=1.25$\\
&&$L_{tst}=0.0$\%&&$L_{tst}=2.5$\%&$L_{tst}=0.0$\%&$L_{tst}=0.0$\%&$L_{tst}=3.5$\%&$L_{tst}=0.0$\%&$L_{tst}=0.0$\%\\
&&$L_{tr}=0.0$\%&&$L_{tr}=2.8$\%&$L_{tr}=0.0$\%&$L_{tr}=0.0$\%&$L_{tr}=4.9$\%&$L_{tr}=0.0$\%&$L_{tr}=0.0$\%\\
&&$N_{sv}$=48&&$N_{sv}$=165&$N_{sv}$=14&$N_{sv}$=21&$N_{sv}$=294&$N_{sv}$=12&$N_{sv}$=4\\\hline
Grindrod &5/7&$f(\bx)=-0.49$&$f(\bx)=-0.99$&&$f(\bx)=0.46$&$f(\bx)=0.39$&$f(\bx)=55.68$&$f(\bx)=7.88$&$f(\bx)=3.87$\\
&&$L_{tst}=0.0$\%&$L_{tst}=2.5$\%&&$L_{tst}=0.0$\%&$L_{tst}=0.0$\%&$L_{tst}=2.0$\%&$L_{tst}=0.0$\%&$L_{tst}=0.0$\%\\
&&$L_{tr}=0.0$\%&$L_{tr}=2.8$\%&&$L_{tr}=0.0$\%&$L_{tr}=0.0$\%&$L_{tr}=1.6$\%&$L_{tr}=0.0$\%&$L_{tr}=0.0$\%\\
&&$N_{sv}$=11&$N_{sv}$=165&&$N_{sv}$=8&$N_{sv}$=9&$N_{sv}$=110&$N_{sv}$=6&$N_{sv}$=3\\\hline
Krapivsky-Bianconi &4/7&$f(\bx)=-3.19$&$f(\bx)=-0.63$&$f(\bx)=-0.46$&&$f(\bx)=0.44$&$f(\bx)=0.10$&$f(\bx)=0.25$&$f(\bx)=0.58$\\
&&$L_{tst}=0.0$\%&$L_{tst}=0.0$\%&$L_{tst}=0.0$\%&&$L_{tst}=6.5$\%&$L_{tst}=0.0$\%&$L_{tst}=0.0$\%&$L_{tst}=0.0$\%\\
&&$L_{tr}=0.0$\%&$L_{tr}=0.0$\%&$L_{tr}=0.0$\%&&$L_{tr}=6.8$\%&$L_{tr}=0.0$\%&$L_{tr}=0.0$\%&$L_{tr}=0.0$\%\\
&&$N_{sv}$=48&$N_{sv}$=14&$N_{sv}$=8&&$N_{sv}$=572&$N_{sv}$=4&$N_{sv}$=9&$N_{sv}$=8\\\hline
Krapivksy &2/7&$f(\bx)=-2.28$&$f(\bx)=-0.06$&$f(\bx)=-0.39$&$f(\bx)=-0.44$&&$f(\bx)=0.23$&$f(\bx)=-0.00$&$f(\bx)=0.15$\\
&&$L_{tst}=0.0$\%&$L_{tst}=0.0$\%&$L_{tst}=0.0$\%&$L_{tst}=6.5$\%&&$L_{tst}=0.0$\%&$L_{tst}=0.0$\%&$L_{tst}=0.0$\%\\
&&$L_{tr}=0.0$\%&$L_{tr}=0.0$\%&$L_{tr}=0.0$\%&$L_{tr}=6.8$\%&&$L_{tr}=0.0$\%&$L_{tr}=0.0$\%&$L_{tr}=0.0$\%\\
&&$N_{sv}$=37&$N_{sv}$=21&$N_{sv}$=9&$N_{sv}$=572&&$N_{sv}$=6&$N_{sv}$=10&$N_{sv}$=8\\\hline
Erdos &2/7&$f(\bx)=-1.18$&$f(\bx)=-16.99$&$f(\bx)=-55.68$&$f(\bx)=-0.10$&$f(\bx)=-0.23$&&$f(\bx)=5.99$&$f(\bx)=1.17$\\
&&$L_{tst}=0.0$\%&$L_{tst}=3.5$\%&$L_{tst}=2.0$\%&$L_{tst}=0.0$\%&$L_{tst}=0.0$\%&&$L_{tst}=0.0$\%&$L_{tst}=0.0$\%\\
&&$L_{tr}=0.0$\%&$L_{tr}=4.9$\%&$L_{tr}=1.6$\%&$L_{tr}=0.0$\%&$L_{tr}=0.0$\%&&$L_{tr}=0.0$\%&$L_{tr}=0.0$\%\\
&&$N_{sv}$=10&$N_{sv}$=294&$N_{sv}$=110&$N_{sv}$=4&$N_{sv}$=6&&$N_{sv}$=6&$N_{sv}$=4\\\hline
Kumar &2/7&$f(\bx)=-0.91$&$f(\bx)=-1.25$&$f(\bx)=-7.88$&$f(\bx)=-0.25$&$f(\bx)=0.00$&$f(\bx)=-5.99$&&$f(\bx)=168.96$\\
&&$L_{tst}=0.0$\%&$L_{tst}=0.0$\%&$L_{tst}=0.0$\%&$L_{tst}=0.0$\%&$L_{tst}=0.0$\%&$L_{tst}=0.0$\%&&$L_{tst}=0.0$\%\\
&&$L_{tr}=0.0$\%&$L_{tr}=0.0$\%&$L_{tr}=0.0$\%&$L_{tr}=0.0$\%&$L_{tr}=0.0$\%&$L_{tr}=0.0$\%&&$L_{tr}=0.0$\%\\
&&$N_{sv}$=8&$N_{sv}$=12&$N_{sv}$=6&$N_{sv}$=9&$N_{sv}$=10&$N_{sv}$=6&&$N_{sv}$=4\\\hline
Vazquez\_K5 &0/7&$f(\bx)=-1.25$&$f(\bx)=-1.25$&$f(\bx)=-3.87$&$f(\bx)=-0.58$&$f(\bx)=-0.15$&$f(\bx)=-1.17$&$f(\bx)=-168.96$&\\
&&$L_{tst}=0.0$\%&$L_{tst}=0.0$\%&$L_{tst}=0.0$\%&$L_{tst}=0.0$\%&$L_{tst}=0.0$\%&$L_{tst}=0.0$\%&$L_{tst}=0.0$\%&\\
&&$L_{tr}=0.0$\%&$L_{tr}=0.0$\%&$L_{tr}=0.0$\%&$L_{tr}=0.0$\%&$L_{tr}=0.0$\%&$L_{tr}=0.0$\%&$L_{tr}=0.0$\%&\\
&&$N_{sv}$=5&$N_{sv}$=4&$N_{sv}$=3&$N_{sv}$=8&$N_{sv}$=8&$N_{sv}$=4&$N_{sv}$=4&\\\hline
\end{tabular}
}
\end{center}\caption{SVM results for C. elegans. $f(x)={\bf w}\cdot \bx_{C. elegans} +b$, $L_{tst}$ is the test loss, $L_{tr}$
the training loss and $N_{sv}$ the number of support vectors.
Results are shown for SVMs trained between every pair of models.
if $f(x)>0$ C. elegans is classified as the row-header, if
$f(x)<0$ as the column-header.}\label{svmceleg}\end{sidewaystable}

\small{
\begin{table}[h]\begin{center}\begin{tabular}{|c|c|c|c|}
\hline RANKING&WORD& $L_{tr}$&ASSOCIATED SUBGRAPHS

\\\hline1&sum U(ATADAA)&0.090\%&\includegraphics[height=15mm]{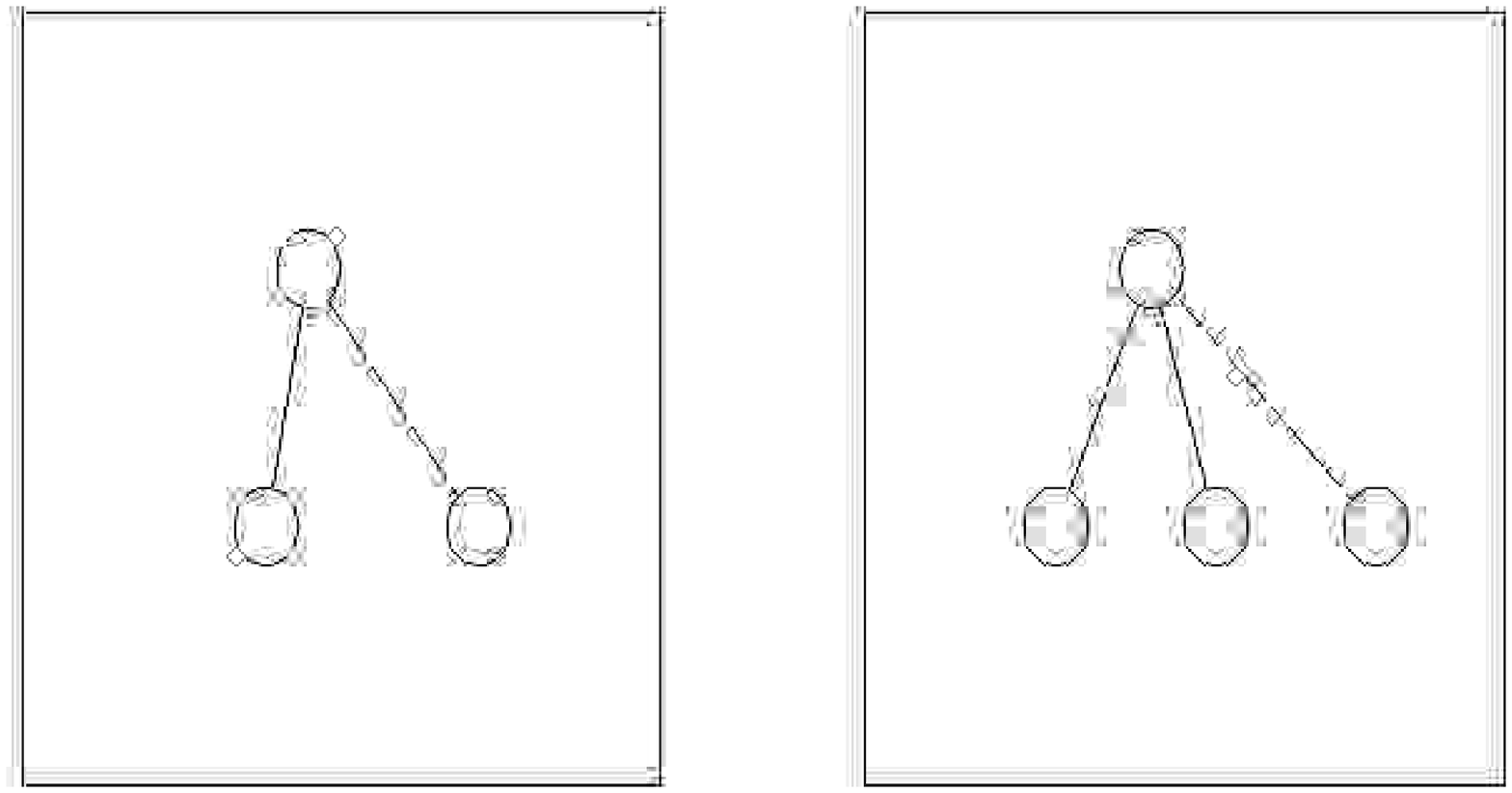}
\\\hline2&nnz D(ATAUAAA)&0.030\%&\includegraphics[height=15mm]{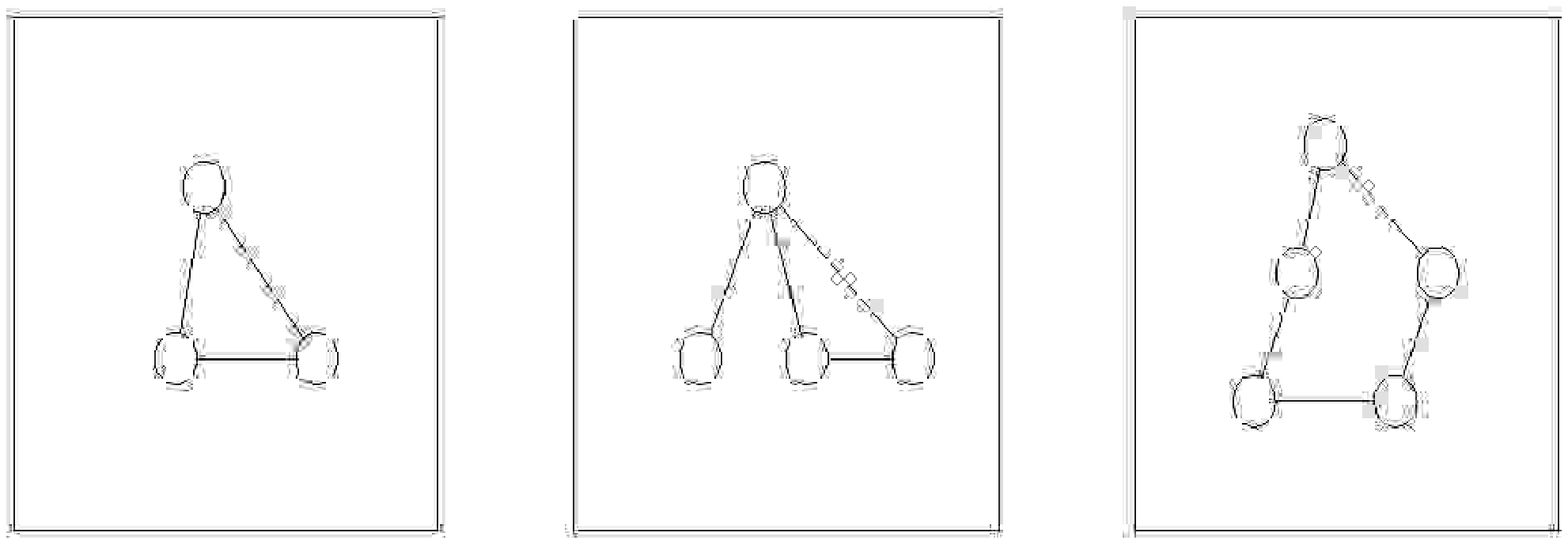}
\\\hline3&nnz D(AA)&0.019\%&\includegraphics[height=15mm]{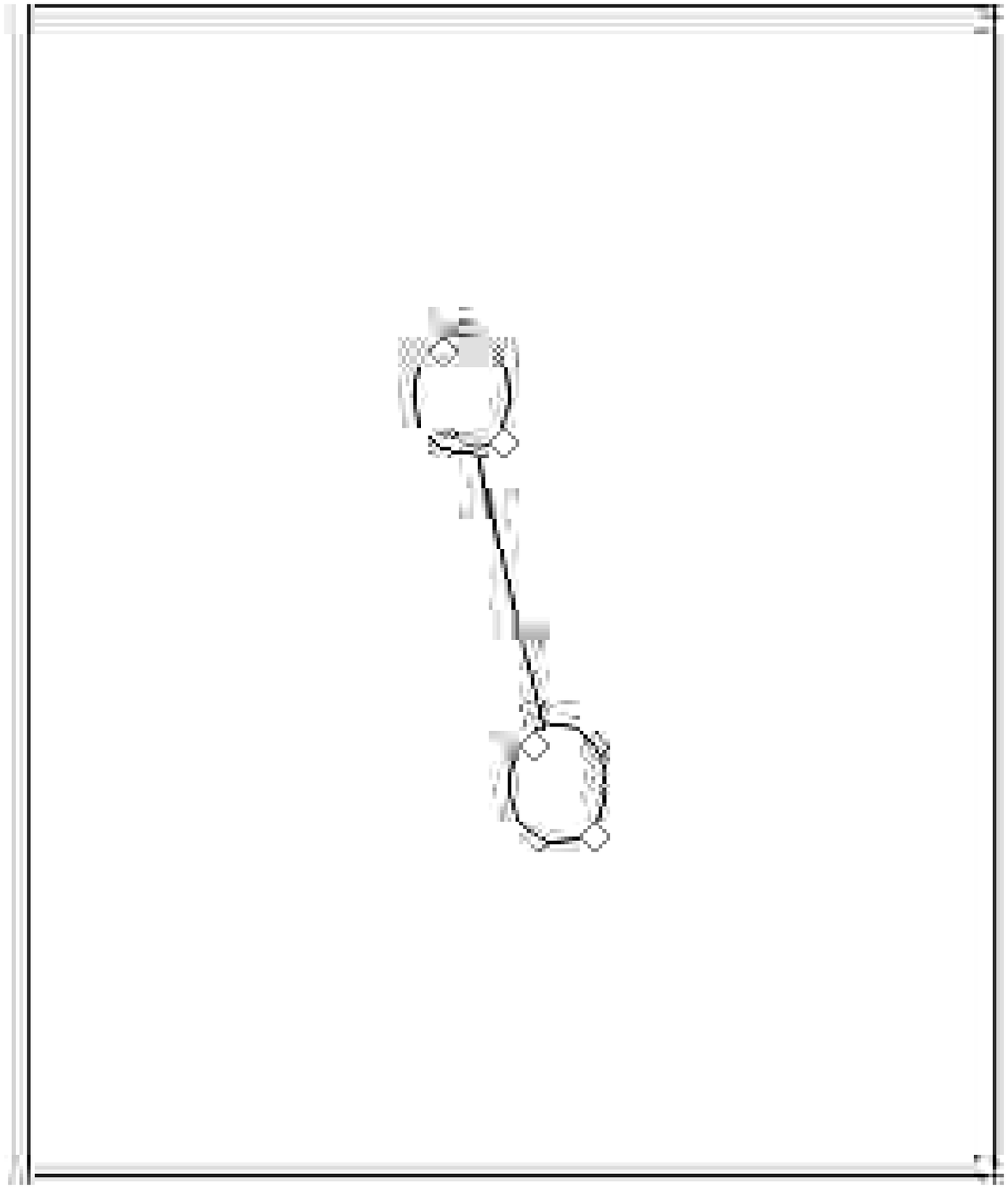}
\\\hline4&nnz(ADATAUAA)&0.016\%&\includegraphics[height=15mm]{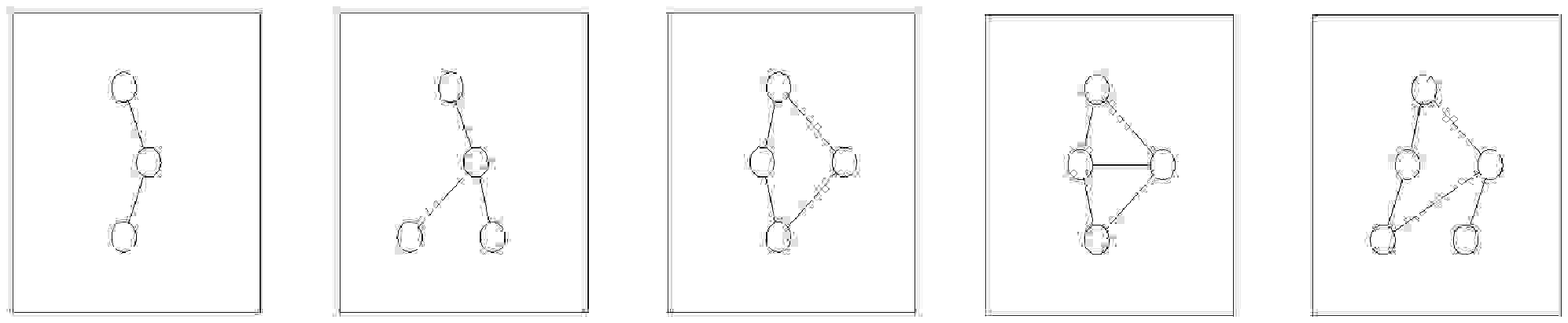}
\\\hline5&nnz(ATAUAUAA)&0.014\%&\includegraphics[height=30mm]{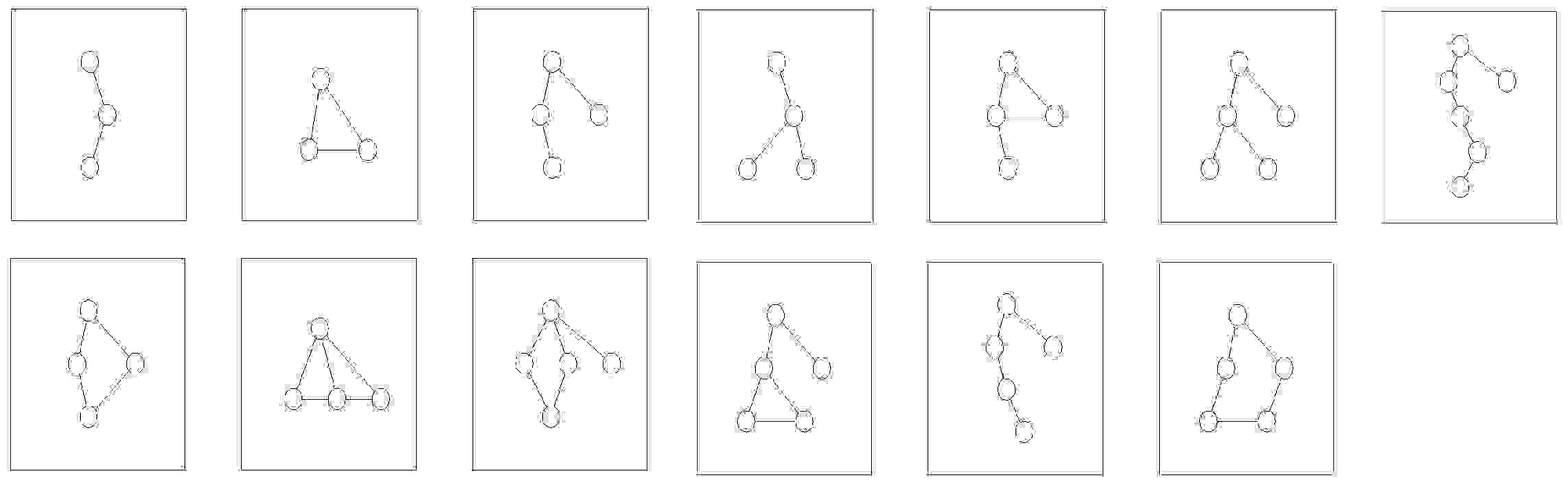}
\\\hline6&sum D(AAUAA)&0.013\%&\includegraphics[height=15mm]{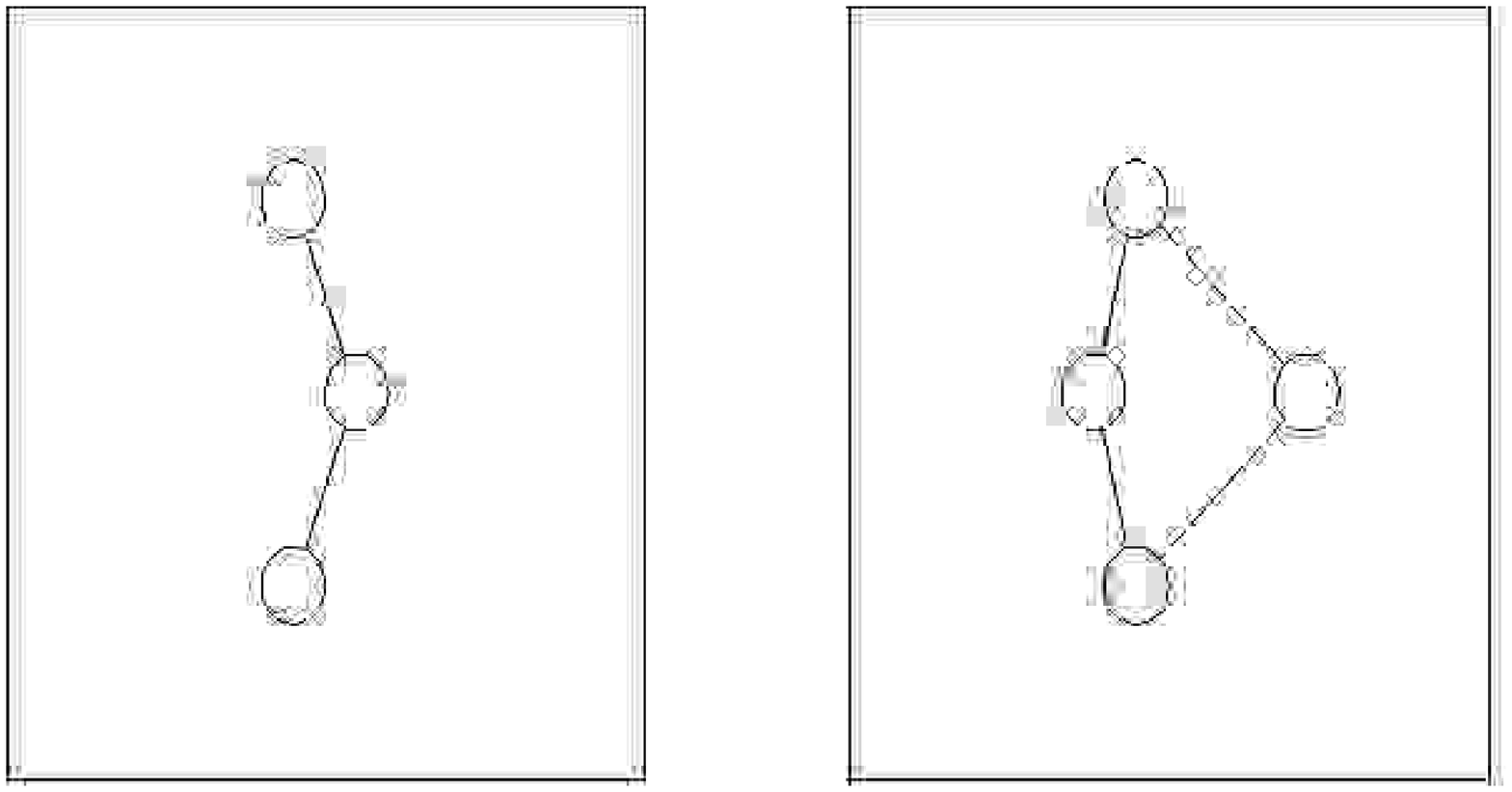}
\\\hline7&nnz D(ATAUAA)&0.013\%&\includegraphics[height=15mm]{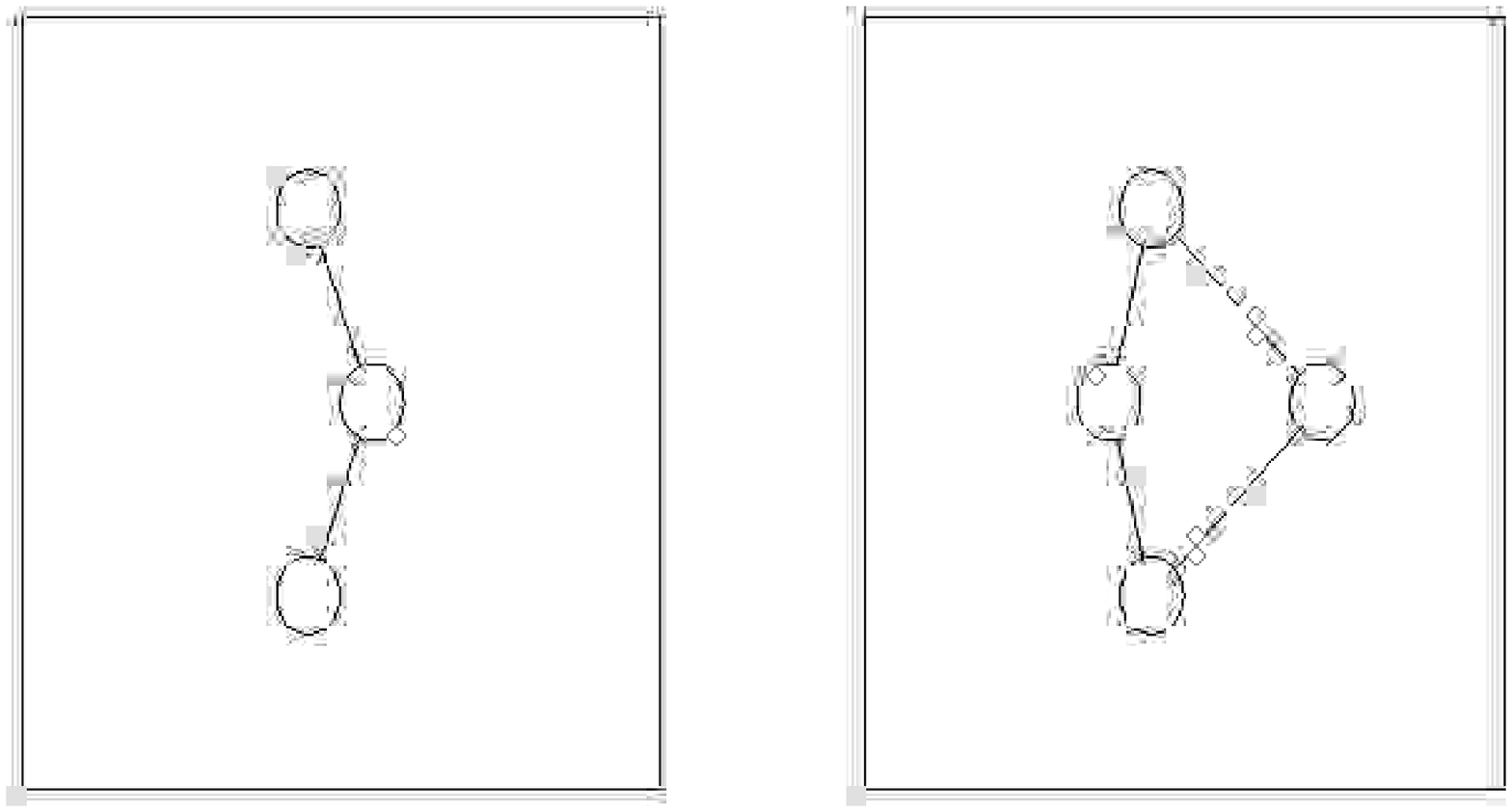}
\\\hline8&sum D(AAAUAA)&0.012\%&\includegraphics[height=15mm]{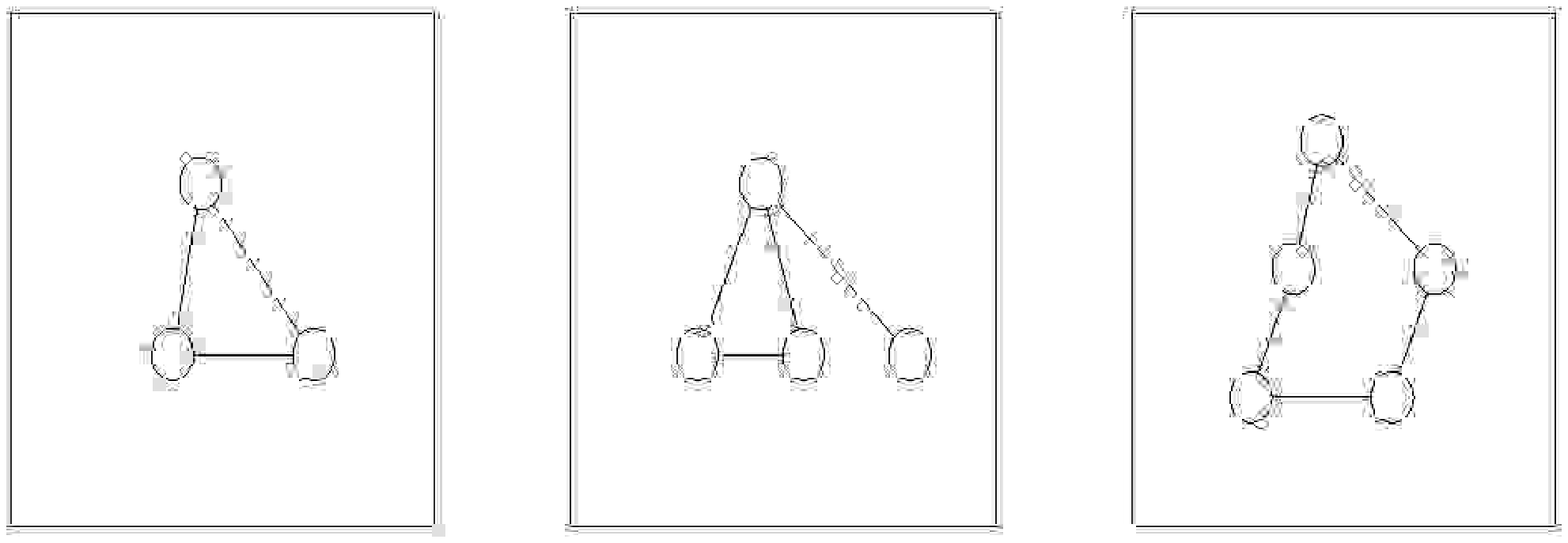}
\\\hline9&nnz D(AAUAA)&0.012\%&\includegraphics[height=15mm]{sumDAAUAA_sym}
\\\hline10&sum(ADAAUAA)&0.012\%&\includegraphics[height=15mm]{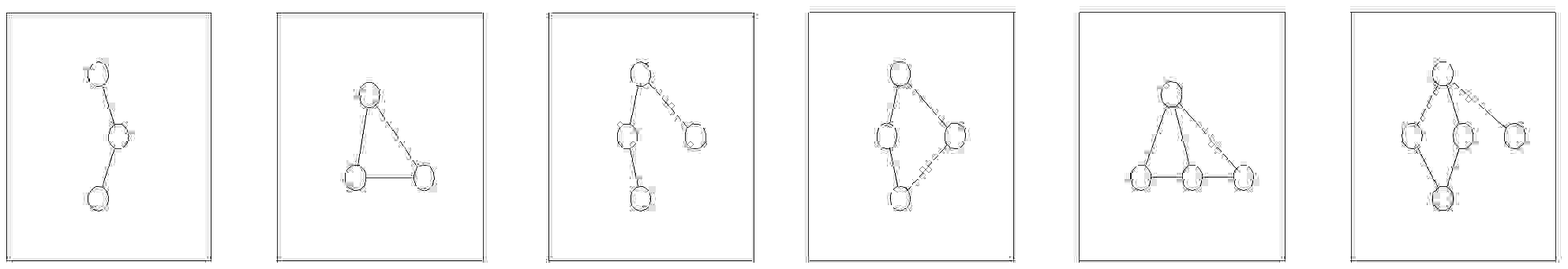}
\\\hline
\end{tabular}\end{center}\caption{
Ranking of words found by binary pairwise trees for the {\bf S.
cerevisiae} training data. $L_{tr}$ for a word ranked $n$ is the
average training loss over all pairwise trees, where every tree
has depth $n$ and splits the data using words $1$ to $n$ in the
given order. 
%See Sec. \ref{sec:wordranking} for details.
}\label{yeastmytree}\end{table} }

\begin{sidewaystable}[h]
\begin{center}
\tiny{
\begin{tabular}
{|l|l|l|l|l|l|l|l|l|l|l|l|l|} \hline
&votes&Sole&Callaway&Flammini&Vazquez&Kim&Grindrod
sym&Barabasi&Erdos&Klemm&Small World&Bianconi\\\hline
Sole &12/10&&$f(\bx)=8.57$&$f(\bx)=4.67$&$f(\bx)=3.67$&$f(\bx)=19.25$&$f(\bx)=10.41$&$f(\bx)=1.75$&$f(\bx)=13.12$&$f(\bx)=4.73$&$f(\bx)=8.56$&$f(\bx)=1.77$\\
&&&$L_{tst}=0.0$\%&$L_{tst}=0.0$\%&$L_{tst}=0.0$\%&$L_{tst}=1.2$\%&$L_{tst}=0.0$\%&$L_{tst}=0.0$\%&$L_{tst}=0.0$\%&$L_{tst}=0.0$\%&$L_{tst}=0.0$\%&$L_{tst}=0.0$\%\\
&&&$L_{tr}=0.0$\%&$L_{tr}=0.0$\%&$L_{tr}=0.0$\%&$L_{tr}=1.2$\%&$L_{tr}=0.0$\%&$L_{tr}=0.0$\%&$L_{tr}=0.0$\%&$L_{tr}=0.0$\%&$L_{tr}=0.0$\%&$L_{tr}=0.0$\%\\
&&&$N_{sv}$=28&$N_{sv}$=36&$N_{sv}$=10&$N_{sv}$=306&$N_{sv}$=20&$N_{sv}$=16&$N_{sv}$=14&$N_{sv}$=41&$N_{sv}$=14&$N_{sv}$=15\\\hline
Callaway &11/10&$f(\bx)=-8.57$&&$f(\bx)=0.27$&$f(\bx)=0.44$&$f(\bx)=0.37$&$f(\bx)=0.76$&$f(\bx)=0.86$&$f(\bx)=0.96$&$f(\bx)=0.57$&$f(\bx)=0.76$&$f(\bx)=0.95$\\
&&$L_{tst}=0.0$\%&&$L_{tst}=0.0$\%&$L_{tst}=0.0$\%&$L_{tst}=0.0$\%&$L_{tst}=0.0$\%&$L_{tst}=0.0$\%&$L_{tst}=0.0$\%&$L_{tst}=0.0$\%&$L_{tst}=0.0$\%&$L_{tst}=0.0$\%\\
&&$L_{tr}=0.0$\%&&$L_{tr}=0.0$\%&$L_{tr}=0.0$\%&$L_{tr}=0.0$\%&$L_{tr}=0.0$\%&$L_{tr}=0.0$\%&$L_{tr}=0.0$\%&$L_{tr}=0.0$\%&$L_{tr}=0.0$\%&$L_{tr}=0.0$\%\\
&&$N_{sv}$=28&&$N_{sv}$=7&$N_{sv}$=2&$N_{sv}$=48&$N_{sv}$=4&$N_{sv}$=10&$N_{sv}$=3&$N_{sv}$=13&$N_{sv}$=4&$N_{sv}$=11\\\hline
Flammini &9/10&$f(\bx)=-4.67$&$f(\bx)=-0.27$&&$f(\bx)=-0.86$&$f(\bx)=0.32$&$f(\bx)=7.52$&$f(\bx)=0.17$&$f(\bx)=0.52$&$f(\bx)=2.38$&$f(\bx)=4.25$&$f(\bx)=0.33$\\
&&$L_{tst}=0.0$\%&$L_{tst}=0.0$\%&&$L_{tst}=0.0$\%&$L_{tst}=0.0$\%&$L_{tst}=6.0$\%&$L_{tst}=0.0$\%&$L_{tst}=0.0$\%&$L_{tst}=0.8$\%&$L_{tst}=7.2$\%&$L_{tst}=0.0$\%\\
&&$L_{tr}=0.0$\%&$L_{tr}=0.0$\%&&$L_{tr}=0.0$\%&$L_{tr}=0.1$\%&$L_{tr}=3.8$\%&$L_{tr}=0.0$\%&$L_{tr}=0.0$\%&$L_{tr}=0.8$\%&$L_{tr}=7.6$\%&$L_{tr}=0.0$\%\\
&&$N_{sv}$=36&$N_{sv}$=7&&$N_{sv}$=29&$N_{sv}$=94&$N_{sv}$=529&$N_{sv}$=10&$N_{sv}$=30&$N_{sv}$=147&$N_{sv}$=384&$N_{sv}$=14\\\hline
Vazquez &9/10&$f(\bx)=-3.67$&$f(\bx)=-0.44$&$f(\bx)=0.86$&&$f(\bx)=0.35$&$f(\bx)=-0.12$&$f(\bx)=0.17$&$f(\bx)=0.95$&$f(\bx)=3.39$&$f(\bx)=0.47$&$f(\bx)=0.23$\\
&&$L_{tst}=0.0$\%&$L_{tst}=0.0$\%&$L_{tst}=0.0$\%&&$L_{tst}=0.0$\%&$L_{tst}=0.0$\%&$L_{tst}=0.0$\%&$L_{tst}=0.0$\%&$L_{tst}=0.0$\%&$L_{tst}=0.0$\%&$L_{tst}=0.0$\%\\
&&$L_{tr}=0.0$\%&$L_{tr}=0.0$\%&$L_{tr}=0.0$\%&&$L_{tr}=0.0$\%&$L_{tr}=0.0$\%&$L_{tr}=0.0$\%&$L_{tr}=0.0$\%&$L_{tr}=0.0$\%&$L_{tr}=0.0$\%&$L_{tr}=0.0$\%\\
&&$N_{sv}$=10&$N_{sv}$=2&$N_{sv}$=29&&$N_{sv}$=20&$N_{sv}$=15&$N_{sv}$=8&$N_{sv}$=7&$N_{sv}$=23&$N_{sv}$=5&$N_{sv}$=8\\\hline
Kim &7/10&$f(\bx)=-19.25$&$f(\bx)=-0.37$&$f(\bx)=-0.32$&$f(\bx)=-0.35$&&$f(\bx)=-1.29$&$f(\bx)=1.41$&$f(\bx)=4.55$&$f(\bx)=1.15$&$f(\bx)=5.60$&$f(\bx)=1.44$\\
&&$L_{tst}=1.2$\%&$L_{tst}=0.0$\%&$L_{tst}=0.0$\%&$L_{tst}=0.0$\%&&$L_{tst}=1.5$\%&$L_{tst}=0.0$\%&$L_{tst}=12.2$\%&$L_{tst}=0.0$\%&$L_{tst}=0.2$\%&$L_{tst}=0.0$\%\\
&&$L_{tr}=1.2$\%&$L_{tr}=0.0$\%&$L_{tr}=0.1$\%&$L_{tr}=0.0$\%&&$L_{tr}=1.4$\%&$L_{tr}=0.0$\%&$L_{tr}=16.7$\%&$L_{tr}=0.0$\%&$L_{tr}=0.4$\%&$L_{tr}=0.0$\%\\
&&$N_{sv}$=306&$N_{sv}$=48&$N_{sv}$=94&$N_{sv}$=20&&$N_{sv}$=107&$N_{sv}$=26&$N_{sv}$=603&$N_{sv}$=55&$N_{sv}$=309&$N_{sv}$=24\\\hline
Grindrod sym &7/10&$f(\bx)=-10.41$&$f(\bx)=-0.76$&$f(\bx)=-7.52$&$f(\bx)=0.12$&$f(\bx)=1.29$&&$f(\bx)=-0.10$&$f(\bx)=3.10$&$f(\bx)=2.75$&$f(\bx)=-2.11$&$f(\bx)=0.08$\\
&&$L_{tst}=0.0$\%&$L_{tst}=0.0$\%&$L_{tst}=6.0$\%&$L_{tst}=0.0$\%&$L_{tst}=1.5$\%&&$L_{tst}=0.0$\%&$L_{tst}=1.2$\%&$L_{tst}=0.0$\%&$L_{tst}=10.5$\%&$L_{tst}=0.0$\%\\
&&$L_{tr}=0.0$\%&$L_{tr}=0.0$\%&$L_{tr}=3.8$\%&$L_{tr}=0.0$\%&$L_{tr}=1.4$\%&&$L_{tr}=0.0$\%&$L_{tr}=0.9$\%&$L_{tr}=0.0$\%&$L_{tr}=10.1$\%&$L_{tr}=0.0$\%\\
&&$N_{sv}$=20&$N_{sv}$=4&$N_{sv}$=529&$N_{sv}$=15&$N_{sv}$=107&&$N_{sv}$=7&$N_{sv}$=66&$N_{sv}$=44&$N_{sv}$=297&$N_{sv}$=11\\\hline
Barabasi &6/10&$f(\bx)=-1.75$&$f(\bx)=-0.86$&$f(\bx)=-0.17$&$f(\bx)=-0.17$&$f(\bx)=-1.41$&$f(\bx)=0.10$&&$f(\bx)=0.17$&$f(\bx)=-2.26$&$f(\bx)=0.37$&$f(\bx)=2.48$\\
&&$L_{tst}=0.0$\%&$L_{tst}=0.0$\%&$L_{tst}=0.0$\%&$L_{tst}=0.0$\%&$L_{tst}=0.0$\%&$L_{tst}=0.0$\%&&$L_{tst}=0.0$\%&$L_{tst}=3.5$\%&$L_{tst}=0.0$\%&$L_{tst}=2.2$\%\\
&&$L_{tr}=0.0$\%&$L_{tr}=0.0$\%&$L_{tr}=0.0$\%&$L_{tr}=0.0$\%&$L_{tr}=0.0$\%&$L_{tr}=0.0$\%&&$L_{tr}=0.0$\%&$L_{tr}=5.6$\%&$L_{tr}=0.0$\%&$L_{tr}=3.0$\%\\
&&$N_{sv}$=16&$N_{sv}$=10&$N_{sv}$=10&$N_{sv}$=8&$N_{sv}$=26&$N_{sv}$=7&&$N_{sv}$=7&$N_{sv}$=281&$N_{sv}$=7&$N_{sv}$=111\\\hline
Erdos &4/10&$f(\bx)=-13.12$&$f(\bx)=-0.96$&$f(\bx)=-0.52$&$f(\bx)=-0.95$&$f(\bx)=-4.55$&$f(\bx)=-3.10$&$f(\bx)=-0.17$&&$f(\bx)=1.35$&$f(\bx)=11.16$&$f(\bx)=0.07$\\
&&$L_{tst}=0.0$\%&$L_{tst}=0.0$\%&$L_{tst}=0.0$\%&$L_{tst}=0.0$\%&$L_{tst}=12.2$\%&$L_{tst}=1.2$\%&$L_{tst}=0.0$\%&&$L_{tst}=0.0$\%&$L_{tst}=0.0$\%&$L_{tst}=0.0$\%\\
&&$L_{tr}=0.0$\%&$L_{tr}=0.0$\%&$L_{tr}=0.0$\%&$L_{tr}=0.0$\%&$L_{tr}=16.7$\%&$L_{tr}=0.9$\%&$L_{tr}=0.0$\%&&$L_{tr}=0.0$\%&$L_{tr}=0.0$\%&$L_{tr}=0.0$\%\\
&&$N_{sv}$=14&$N_{sv}$=3&$N_{sv}$=30&$N_{sv}$=7&$N_{sv}$=603&$N_{sv}$=66&$N_{sv}$=7&&$N_{sv}$=24&$N_{sv}$=10&$N_{sv}$=9\\\hline
Klemm &2/10&$f(\bx)=-4.73$&$f(\bx)=-0.57$&$f(\bx)=-2.38$&$f(\bx)=-3.39$&$f(\bx)=-1.15$&$f(\bx)=-2.75$&$f(\bx)=2.26$&$f(\bx)=-1.35$&&$f(\bx)=-4.53$&$f(\bx)=2.14$\\
&&$L_{tst}=0.0$\%&$L_{tst}=0.0$\%&$L_{tst}=0.8$\%&$L_{tst}=0.0$\%&$L_{tst}=0.0$\%&$L_{tst}=0.0$\%&$L_{tst}=3.5$\%&$L_{tst}=0.0$\%&&$L_{tst}=0.0$\%&$L_{tst}=1.8$\%\\
&&$L_{tr}=0.0$\%&$L_{tr}=0.0$\%&$L_{tr}=0.8$\%&$L_{tr}=0.0$\%&$L_{tr}=0.0$\%&$L_{tr}=0.0$\%&$L_{tr}=5.6$\%&$L_{tr}=0.0$\%&&$L_{tr}=0.0$\%&$L_{tr}=0.9$\%\\
&&$N_{sv}$=41&$N_{sv}$=13&$N_{sv}$=147&$N_{sv}$=23&$N_{sv}$=55&$N_{sv}$=44&$N_{sv}$=281&$N_{sv}$=24&&$N_{sv}$=33&$N_{sv}$=106\\\hline
Small World &2/10&$f(\bx)=-8.56$&$f(\bx)=-0.76$&$f(\bx)=-4.25$&$f(\bx)=-0.47$&$f(\bx)=-5.60$&$f(\bx)=2.11$&$f(\bx)=-0.37$&$f(\bx)=-11.16$&$f(\bx)=4.53$&&$f(\bx)=-0.02$\\
&&$L_{tst}=0.0$\%&$L_{tst}=0.0$\%&$L_{tst}=7.2$\%&$L_{tst}=0.0$\%&$L_{tst}=0.2$\%&$L_{tst}=10.5$\%&$L_{tst}=0.0$\%&$L_{tst}=0.0$\%&$L_{tst}=0.0$\%&&$L_{tst}=0.0$\%\\
&&$L_{tr}=0.0$\%&$L_{tr}=0.0$\%&$L_{tr}=7.6$\%&$L_{tr}=0.0$\%&$L_{tr}=0.4$\%&$L_{tr}=10.1$\%&$L_{tr}=0.0$\%&$L_{tr}=0.0$\%&$L_{tr}=0.0$\%&&$L_{tr}=0.0$\%\\
&&$N_{sv}$=14&$N_{sv}$=4&$N_{sv}$=384&$N_{sv}$=5&$N_{sv}$=309&$N_{sv}$=297&$N_{sv}$=7&$N_{sv}$=10&$N_{sv}$=33&&$N_{sv}$=9\\\hline
Bianconi &1/10&$f(\bx)=-1.77$&$f(\bx)=-0.95$&$f(\bx)=-0.33$&$f(\bx)=-0.23$&$f(\bx)=-1.44$&$f(\bx)=-0.08$&$f(\bx)=-2.48$&$f(\bx)=-0.07$&$f(\bx)=-2.14$&$f(\bx)=0.02$&\\
&&$L_{tst}=0.0$\%&$L_{tst}=0.0$\%&$L_{tst}=0.0$\%&$L_{tst}=0.0$\%&$L_{tst}=0.0$\%&$L_{tst}=0.0$\%&$L_{tst}=2.2$\%&$L_{tst}=0.0$\%&$L_{tst}=1.8$\%&$L_{tst}=0.0$\%&\\
&&$L_{tr}=0.0$\%&$L_{tr}=0.0$\%&$L_{tr}=0.0$\%&$L_{tr}=0.0$\%&$L_{tr}=0.0$\%&$L_{tr}=0.0$\%&$L_{tr}=3.0$\%&$L_{tr}=0.0$\%&$L_{tr}=0.9$\%&$L_{tr}=0.0$\%&\\
&&$N_{sv}$=15&$N_{sv}$=11&$N_{sv}$=14&$N_{sv}$=8&$N_{sv}$=24&$N_{sv}$=11&$N_{sv}$=111&$N_{sv}$=9&$N_{sv}$=106&$N_{sv}$=9&\\\hline

\end{tabular}
}
\end{center}
\caption{SVM results for S. cerevisiae (only 11 models out of 13
shown). $f(x)={\bf w}\cdot \bx_{S. cerevisiae} +b$, $L_{tst}$ is
the test loss, $L_{tr}$ the training loss and $N_{sv}$ the number
of support vectors. Results are shown for SVMs trained between
every pair of models. if $f(x)>0$ S. cerevisiae is classified as
the row-header, if $f(x)<0$ as the column-header.}
\label{svmyeast}
\end{sidewaystable}

\begin{sidewaystable}[h]
\begin{center}
\tiny{
\begin{tabular}{|l|l|l|l|l|l|l|l|l|l|l|l|} \hline
&Sole&Callaway&Flammini&Vazquez&Kim&Grindrod
sym&Barabasi&Erdos&Klemm&Small World&Bianconi\\\hline
Sole &&nnz(AAAAA)&nnz(AAAAA)&nnz(AA)&nnz(ADA)&nnz(AA)&nnz D(AA)&nnz(AA)&nnz D(AA)&nnz(AA)&nnz D(AA)\\
&&$L_{tst}=0.3$\%&$L_{tst}=3.8$\%&$L_{tst}=0.0$\%&$L_{tst}=7.2$\%&$L_{tst}=0.0$\%&$L_{tst}=0.0$\%&$L_{tst}=0.0$\%&$L_{tst}=0.0$\%&$L_{tst}=0.0$\%&$L_{tst}=0.0$\%\\
&&$L_{tr}=0.5$\%&$L_{tr}=2.5$\%&$L_{tr}=0.0$\%&$L_{tr}=5.2$\%&$L_{tr}=0.0$\%&$L_{tr}=0.0$\%&$L_{tr}=0.0$\%&$L_{tr}=0.0$\%&$L_{tr}=0.0$\%&$L_{tr}=0.0$\%\\\hline
Callaway &&&nnz(AAAAA)&nnz(AA)&nnz(ADATAUAA)&nnz(AA)&nnz(AA)&nnz(AA)&nnz D(AA)&nnz(AA)&nnz(AA)\\
&&&$L_{tst}=0.0$\%&$L_{tst}=0.0$\%&$L_{tst}=3.0$\%&$L_{tst}=0.0$\%&$L_{tst}=0.0$\%&$L_{tst}=0.0$\%&$L_{tst}=0.0$\%&$L_{tst}=0.0$\%&$L_{tst}=0.0$\%\\
&&&$L_{tr}=0.0$\%&$L_{tr}=0.0$\%&$L_{tr}=5.2$\%&$L_{tr}=0.0$\%&$L_{tr}=0.0$\%&$L_{tr}=0.0$\%&$L_{tr}=0.0$\%&$L_{tr}=0.0$\%&$L_{tr}=0.0$\%\\\hline
Flammini &&&&nnz D(AAUAA)&nnz D(AAA)&nnz U(ATADAAA)&nnz D(AAUAA)&sum(ADAAA)&nnz D(AAUAA)&sum D(AAUAA)&nnz(AAA)\\
&&&&$L_{tst}=0.0$\%&$L_{tst}=13.8$\%&$L_{tst}=14.0$\%&$L_{tst}=0.0$\%&$L_{tst}=0.0$\%&$L_{tst}=0.5$\%&$L_{tst}=8.5$\%&$L_{tst}=0.0$\%\\
&&&&$L_{tr}=0.1$\%&$L_{tr}=11.1$\%&$L_{tr}=13.4$\%&$L_{tr}=0.2$\%&$L_{tr}=0.1$\%&$L_{tr}=0.2$\%&$L_{tr}=8.9$\%&$L_{tr}=0.0$\%\\\hline
Vazquez &&&&&nnz D(AA)&nnz D(ATAUAA)&nnz(AA)&nnz D(AA)&sum D(AAA)&nnz(AA)&nnz(AA)\\
&&&&&$L_{tst}=0.0$\%&$L_{tst}=0.0$\%&$L_{tst}=0.0$\%&$L_{tst}=0.0$\%&$L_{tst}=0.0$\%&$L_{tst}=0.0$\%&$L_{tst}=0.0$\%\\
&&&&&$L_{tr}=0.0$\%&$L_{tr}=0.0$\%&$L_{tr}=0.0$\%&$L_{tr}=0.0$\%&$L_{tr}=0.0$\%&$L_{tr}=0.0$\%&$L_{tr}=0.0$\%\\\hline
Kim &&&&&&nnz(AAAAA)&nnz D(AA)&sum U(ATADAA)&nnz D(AA)&nnz(AA)&nnz D(AA)\\
&&&&&&$L_{tst}=0.8$\%&$L_{tst}=0.0$\%&$L_{tst}=8.0$\%&$L_{tst}=0.0$\%&$L_{tst}=0.0$\%&$L_{tst}=0.0$\%\\
&&&&&&$L_{tr}=0.6$\%&$L_{tr}=0.0$\%&$L_{tr}=9.2$\%&$L_{tr}=0.0$\%&$L_{tr}=0.0$\%&$L_{tr}=0.0$\%\\\hline
Grindrod sym &&&&&&&nnz(AA)&nnz D(ATAUAAA)&nnz(ADATAUAA)&nnz D(ATAUAAA)&nnz(AA)\\
&&&&&&&$L_{tst}=0.0$\%&$L_{tst}=0.3$\%&$L_{tst}=0.0$\%&$L_{tst}=18.5$\%&$L_{tst}=0.0$\%\\
&&&&&&&$L_{tr}=0.0$\%&$L_{tr}=0.1$\%&$L_{tr}=0.0$\%&$L_{tr}=19.2$\%&$L_{tr}=0.0$\%\\\hline
Barabasi &&&&&&&&nnz(AA)&nnz D(ATAUAA)&nnz(AA)&nnz D(ATAUAA)\\
&&&&&&&&$L_{tst}=0.0$\%&$L_{tst}=2.0$\%&$L_{tst}=0.0$\%&$L_{tst}=0.0$\%\\
&&&&&&&&$L_{tr}=0.0$\%&$L_{tr}=2.7$\%&$L_{tr}=0.0$\%&$L_{tr}=0.0$\%\\\hline
Erdos &&&&&&&&&nnz D(AA)&nnz(AA)&nnz(AA)\\
&&&&&&&&&$L_{tst}=0.0$\%&$L_{tst}=0.0$\%&$L_{tst}=0.0$\%\\
&&&&&&&&&$L_{tr}=0.0$\%&$L_{tr}=0.0$\%&$L_{tr}=0.0$\%\\\hline
Klemm &&&&&&&&&&nnz D(AA)&nnz D(ATAUAA)\\
&&&&&&&&&&$L_{tst}=0.0$\%&$L_{tst}=0.0$\%\\
&&&&&&&&&&$L_{tr}=0.0$\%&$L_{tr}=0.0$\%\\\hline
Small World &&&&&&&&&&&nnz(AA)\\
&&&&&&&&&&&$L_{tst}=0.0$\%\\
&&&&&&&&&&&$L_{tr}=0.0$\%\\\hline
Bianconi &&&&&&&&&&&\\
&&&&&&&&&&&\\
&&&&&&&&&&&\\\hline
\end{tabular}
}
\end{center}
\caption{ Most discriminative words for the S. cerevisiae
training data based on lowest test loss by 1-dimensional splitting
for every pair of models. $L_{tst}$ is the test loss and $L_{tr}$
the training loss.} \label{yeastsingle}
\end{sidewaystable}

%\end{widetext}
}
\end{document}
 \end